\documentclass[12pt]{article}
\usepackage{amsmath, amsfonts, graphicx, graphics, bm, float, adjustbox, lscape, threeparttable, rotating, makecell, setspace, tabularx, multirow, booktabs, natbib, caption, subcaption, chngcntr, dsfont}
\usepackage[table,xcdraw]{xcolor}

\usepackage[utf8]{inputenc}
\usepackage[top=3cm,left=3cm,right=3cm,bottom=3cm]{geometry}
\usepackage[graphicx]{realboxes}
\usepackage [english]{babel}
\usepackage [autostyle, english = american]{csquotes}
\usepackage[all,cmtip]{xy} 
\MakeOuterQuote{"}
\usepackage[toc,page]{appendix}

\usepackage[T1]{fontenc}

\usepackage{hyperref}
\hypersetup{
colorlinks=true,
citecolor=blue,
linkcolor=blue,
filecolor=blue,      
urlcolor=blue,
}

\graphicspath{{pictures}}

\begin{document}

\begin{titlepage}
\title{\Large One Call Away. Ownership Chains and Ease of Communication in Multinational Enterprises \thanks{\scriptsize{We would like to thank Lionel Fontagné, Thierry Mayer, Ariell Reshef, and Farid Toubal for useful comments. We are also grateful to participants at the Royal Economic Society Annual Conference (Royal Economic Society) in 2024, the ASSA Annual Meeting in 2024, the European Trade Study Group (ETSG) in 2023, the Annual Conference on the Economics of Global Interactions in 2023, the Italian Trade Study Group in 2023, and the European Trade Study Group in 2022. Armando Rungi claims financial support from the PRIN project 2022W2245M, financed by the European Union - Next Generation EU.}}}
\author{Stefania Miricola\thanks{\scriptsize Mail to:stefania.miricola@ec.europa.eu. Joint Research Centre - Ispra, Italy.} \and Armando Rungi\thanks{\scriptsize Mail to: armando.rungi@imtlucca.it. Laboratory on Firms Markets Technologies (FMT), IMT School for Advanced Studies, Piazza San Francesco 19 - 55100 Lucca, Italy.} \and Gianluca Santoni \thanks{\scriptsize Mail to: gianluca.santoni@psemail.eu. Paris School of Economics - Paris, France.}} 
 \date{}
\maketitle
\vspace{0.1in}
\begin{abstract}
\footnotesize
\noindent  
This study examines how multinational enterprises structure ownership chains to coordinate subsidiaries across multiple national borders. Using a unique global dataset, we first document key stylized facts: 54\% of subsidiaries are controlled through indirect ownership, and ownership chains can span up to seven countries. In particular, we find that subsidiaries further down the control hierarchy tend to be more geographically distant from the parent and operate in different time zones. This suggests that the ease of communication along ownership chains is a critical determinant of their structure. On the other hand, tax optimization strategies are not correlated with locations along ownership chains. Motivated by previous findings, we develop a location choice model in which parent firms compete for corporate control of final subsidiaries, but monitoring is costly, and they can delegate control to an intermediate affiliate in another jurisdiction. The model generates a two-stage empirical strategy: (i) a trilateral equation that determines the location of an intermediate affiliate conditional on the location of final subsidiaries; and (ii) a bilateral equation that predicts the location of final investment. Our empirical estimates confirm that the ease of communication at the country level has a significant influence on the location decisions of affiliates along ownership chains. Our findings underscore the importance of communication frictions in shaping global corporate structures, and provide new insights into the geography of multinational ownership networks.\\

\vspace{0cm}
\noindent\textbf{Keywords:} multinational firms; ownership structure; corporate control; ease of communication; knowledge hierarchies\\ 
\noindent\textbf{JEL Codes:}  F23, L22, L23, G34 \\
\bigskip
\end{abstract}
\setcounter{page}{0}
\thispagestyle{empty}
\end{titlepage}
\pagebreak \newpage

\onehalfspacing

\maketitle

\section{Introduction}

Multinational enterprises (MNEs) organize their global operations through complex ownership networks, which function as management conduits for subsidiaries across multiple countries. These structures often involve multi-tiered chains of intermediate affiliates, extending well beyond direct parent-subsidiary links. Recent evidence indicates that up to 56\% of subsidiaries worldwide are indirectly controlled through an intermediate entity rather than by their immediate parent company \citep{unctad2016inv}.\\

\noindent Despite their prevalence, the strategic rationale behind indirect ownership structures and their role in investment decisions remain underexplored. What drives MNEs' decisions on the location of intermediate affiliates along these chains? To what extent do geographic distance, communication frictions, and control mechanisms influence the organization of ownership networks?\\

\noindent This paper addresses these questions by combining new empirical evidence on the prevalence and structure of indirect ownership chains with a structural model explaining their formation. In doing so, we analyze a global database of 226,931 multinational parents that control 1,785,493 subsidiaries in 190 countries as of 2019. We document new evidence on the geography and organization of indirect ownership chains. Motivated by this evidence, we propose a location choice model to explain the emergence of these structures. Our model is based on the premise that effective corporate governance requires efficient communication between parent and subsidiary managers. As a result, MNEs incur delegation costs when they use intermediate affiliates in strategically located jurisdictions to monitor distant subsidiaries.\\

\noindent The theoretical model provides a formal framework for understanding the geographic and hierarchical organization of ownership chains. Specifically, we estimate a system of two simultaneous equations: i) one equation that models the choice of intermediate affiliate locations, given the placement of newly acquired subsidiaries; and ii) another equation that explains the geographic distribution of newly acquired subsidiaries, incorporating factors that influence firms' control and coordination costs. Our results indicate that communication frictions play a fundamental role in determining the structure of ownership chains. The model's predictions remain robust when tested against alternative explanations, including tax optimization, labor market comparative advantage, and cultural proximity.  \\

\noindent In our dataset on global parent-subsidiary relationships in 2019, indirect control emerges as the dominant strategy, with 54\% of subsidiaries worldwide controlled by intermediate affiliates.\footnote{This figure closely matches the 56\% reported by \cite{unctad2016inv}.} Complex MNEs - those that develop chains of ownership - account for 95\% of global sales in our sample, underscoring the economic importance of this organizational structure. Exploring firm heterogeneity, we find that large MNEs with more than 100 subsidiaries account for only 1.2\% of the global sample, yet they operate in over 19 countries and span more than 15 industries, highlighting their outsized international footprint. Building on this heterogeneity, we examine the hierarchical depth and geographic structure of MNEs' ownership chains. Our findings show that hierarchical organization is indeed a defining characteristic of global firms. Nearly 80\% of subsidiaries are located within three hierarchical levels of the parent, implying that two intermediate affiliates typically mediate between the parent and the ultimate subsidiary. In a few cases, however, ownership chains extend over more than twelve levels, illustrating the remarkable depth of some corporate networks. \\

\noindent Prior research documented that complex legal and organizational structures of multinational enterprises are used to optimize taxation and, thus, shift profits from one jurisdiction to another \citep{francois2025tax}. On the one hand, we do not find significant patterns in the location of subsidiaries along the ownership chains and their tax burden. On the other hand, the location in tax havens or the establishment of empty shells, such as Special Purpose Entities, do not explain the sequence of the ownership chain.\\

\noindent Notably, we find a strong correlation between the hierarchical position of subsidiaries and geographic distance. Subsidiaries further down the ownership chain tend to be further away from their parent companies. This suggests that the ease of communication plays a role in shaping the ownership structure. To measure this, we analyze the overlap of working hours between countries along the chain, using time zone differences as a proxy for communication frictions, following \cite{stein2007longitude} and \cite{bahar2020hardships}. In particular, our results show that subsidiaries that are located deeper in the hierarchy have progressively less overlap in working hours with the managers of the parent company. Meanwhile, subsidiaries closer to the parent in the hierarchy experience more synchronous work schedules, which facilitate direct managerial supervision and coordination. \\

\noindent Building on previous empirical evidence, we develop a structural model in which ownership chains emerge from a two-step investment decision. First, parent firms compete in a global auction for corporate control of new subsidiaries. Their bids take into account the option of delegating monitoring of the subsidiary to an intermediate affiliate located in a strategically chosen third country. The core of our model is an inspection game in which corporate control generates value based on the productivity of the parent firm and its network on the one hand, and the performance of newly acquired subsidiaries on the other. The valuation of new subsidiaries depends on managerial effort and monitoring intensity, both of which entail inspection costs. Importantly, these costs include the option of delegating monitoring to an intermediate subsidiary, which creates a trade-off between direct control and delegated monitoring. Second, once the intermediate subsidiary is established, the parent firm makes the final investment decision regarding the placement of the subsidiary. \\

\noindent Our first testable equation estimates the likelihood that a parent firm in an origin country chooses a particular intermediate jurisdiction to establish an affiliate that controls a subsidiary in a destination country. Empirically, this corresponds to an aggregate tripartite model, where the dependent variable is the share of origin-intermediate-destination triplets in total origin-destination investment decisions. On the right-hand side, we proxy for delegation and monitoring costs by including frictions between origin and intermediate countries, as well as between intermediate and destination countries. Following our initial evidence, we include time zone differences as a proxy for communication frictions and further control for geographic distance, common language, colonial ties, common legal origins, home bias, and regional trade agreements. The second testable equation examines the probability of gaining global market control for subsidiaries in a destination country. Here, the dependent variable is the number of subsidiaries controlled by firms from a source country in a given destination. The right-hand side includes the predicted inspection costs between the source and destination, obtained from the first equation. Together, these two equations form a system of equations, where the estimates from the first equation serve as inputs to solve the second equation, capturing the interdependence between delegation decisions and final investment decisions.\\

\noindent Our empirical analysis confirms our predictions regarding inspection costs, with ease of communication playing an essential role. When we include estimated inspection costs in the second equation, our model successfully predicts investment decisions between source and destination countries.\\

\noindent We further challenge our results through a series of robustness checks. First, we address the concern that alternative explanations may drive the organization of ownership chains. To test this, we include additional controls for tax differentials, wage differentials, and cultural/language proximity. Our results show mixed evidence for tax and wage effects, while cultural proximity consistently influences the selection of both intermediate and final subsidiary locations. In all cases, however, communication frictions remain a key determinant of ownership structure. We also examine whether our results are driven by sample composition effects. To test this, we conduct separate analyses for manufacturing, services, and financial groups. Across all subsamples, our core findings remain robust, confirming that the observed patterns in ownership chains are not driven by industry-specific characteristics.\\

\begin{figure}[H]
\centering
\caption{Across the oceans - the corporate tree of QPS Holding}
\label{fig:QPS_holding}
\includegraphics[width=0.7\linewidth]{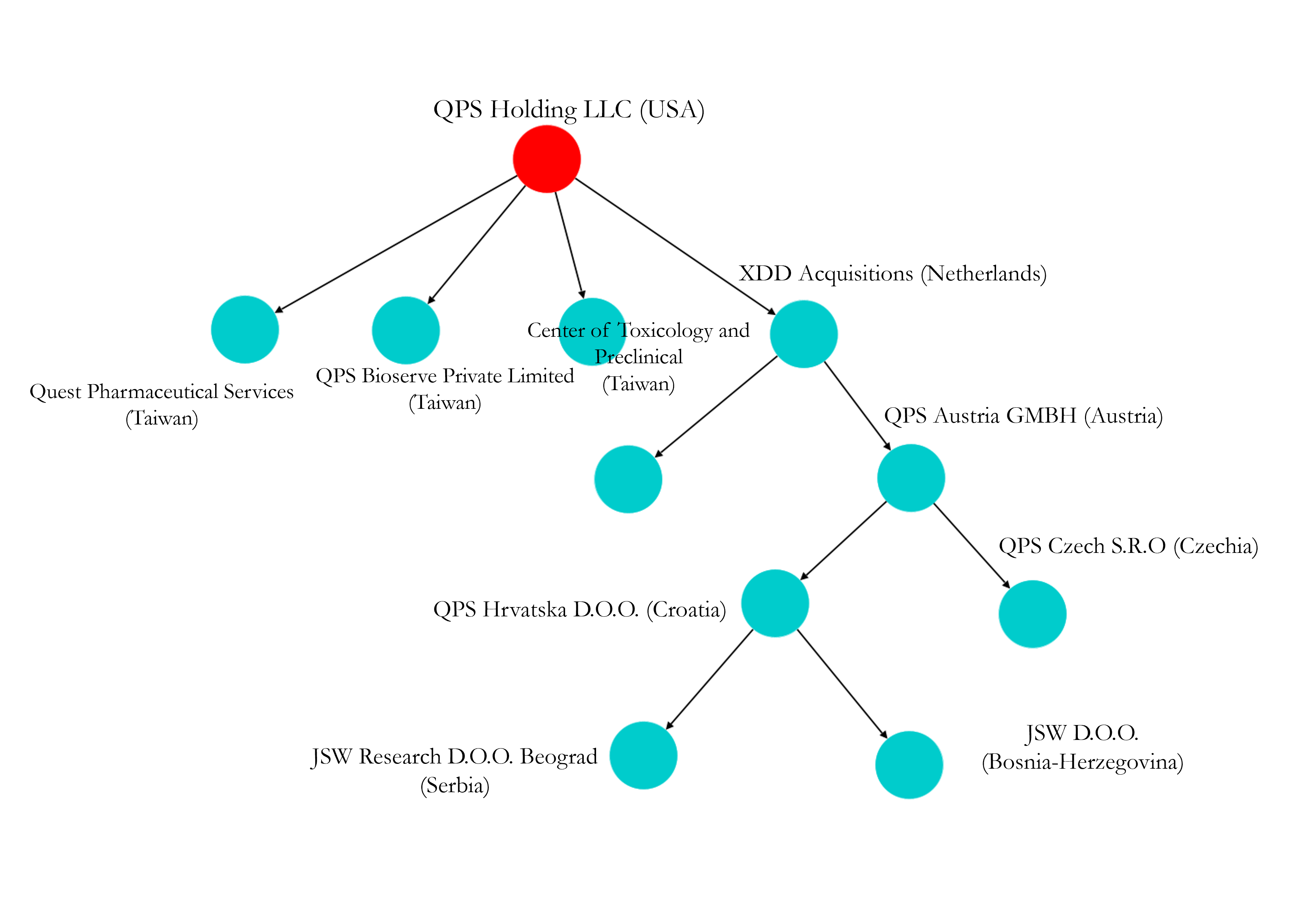}
\begin{tablenotes}
\footnotesize
\item Note: The figure shows the corporate structure of an MNE extracted from our sample. The node on top represents the parent company, other downstream nodes represent subsidiaries, and arrows represent ownership linkages directed from one company to another. The parent company in the USA coordinates a chain across the Atlantic Ocean with an evident orientation from West to East, while directly owned subsidiaries on the other side of the Pacific Ocean are located in Taiwan. 
\end{tablenotes}
\end{figure}

\noindent To illustrate the structure of ownership chains visually, we present the case of a medium-sized MNE from our data in Figure \ref{fig:QPS_holding}. QPS Holding has a corporate structure that spans three continents: the Americas, Asia, and Europe. QPS Holding is an American multinational active in bioanalytical, preclinical, and clinical research services. The parent company is at the top of the ownership hierarchy, with direct or indirect ownership of subsidiaries through four levels of hierarchy. For example, \textit{JSW Research DOO Beograd}, located in Serbia, is owned by a chain that includes three intermediate subsidiaries: \textit{XDD Acquisition B.V.}, \textit{QPS Austria GMBH} and \textit{QPS Hrvatska DOO}. This case illustrates the multi-tiered organization of multinational corporations that we describe throughout the paper, and the role of intermediate subsidiaries in cross-border corporate structures.

\noindent Starting from its headquarters, this chain of ownership crosses four national borders - the United States, the Netherlands, Austria, and Croatia - before reaching its final subsidiary in Serbia. Notably, the chain follows a geographic trajectory that extends across the Atlantic Ocean, with affiliates located progressively farther from the United States. In the following sections, we provide evidence that this relationship between hierarchical position and geographic distance is statistically significant when analyzing the entire dataset.\\ 

\noindent  The rest of the paper is organized as follows. Section \ref{sec: lit} reviews the relevant literature. Section \ref{sec: data} introduces the data set and presents stylized facts on the heterogeneity, economic relevance, and geographic distribution of ownership chains. Section \ref{sec: model} outlines our structural model, while Section \ref{sec: str} describes the empirical strategy. The results are presented in Section \ref{sec: res} and further verified by robustness checks in Section \ref{sec: RC}. Finally, Section \ref{sec: concl} provides concluding remarks.

\section{Literature review}
\label{sec: lit}

A central question in the study of the network of multinational enterprises (MNEs) is why firms establish complex hierarchical ownership structures. Previous economic literature has identified several drivers, including: knowledge transfer, governance and control, tax optimization, and production optimization. These factors interact to shape ownership chains, as firms must balance efficiency, control, and regulatory arbitrage when structuring their global operations.
\noindent This paper makes two contributions to the literature. First, we document the widespread presence of hierarchical ownership structures in MNEs' networks, highlighting their geography and organizational complexity. Second, we examine how communication frictions - resulting from geographical, institutional, and contractual barriers - contribute to the formation of such structures. We argue that beyond tax motives and production efficiency, the need to maintain control, monitor subsidiaries, and reduce coordination costs plays a crucial role in shaping ownership chains.\\

\noindent One of the fundamental explanations for the emergence of hierarchical structures is knowledge transmission and management efficiency. \cite{garicano2000hierarchies} conceptualizes firms as knowledge-based hierarchies in which managers' hierarchical differentiation optimizes knowledge allocation. Routine problems are handled at lower levels, while complex problems are escalated. This framework has been extended to multinational firms, emphasizing the importance of structuring the organization to facilitate knowledge flows across geographically dispersed units \citep{antras2006offshoring, caliendo2012impact}.
\noindent \cite{gumpert2018organization} shows that firms face heterogeneous communication costs when transferring knowledge from headquarters to foreign affiliates, reinforcing the need for hierarchical complexity. Other authors have used a similar theoretical framework to study domestic and multinational business groups. \cite{altomonte2026business} propose and test a model in which problem-solving efficiency helps determine the optimal hierarchical form of a corporate group, with parent firms supervising subsidiaries and managing communication among them. \cite{altomonte2013business} examine the relationship between hierarchical complexity and vertical integration, and find consistencies with predictions from knowledge-based models.\\

\noindent However, knowledge transfer alone does not fully explain organizational complexity. Beyond efficiency gains, firms must also retain control over dispersed operations, especially when facing institutional and contractual frictions. \cite{gumpert2022firm} show that when the CEO’s ability to oversee operations is limited, firms introduce intermediate layers of management to optimize control. Similarly, \cite{aghion1997formal} highlight the trade-off between corporate control and decentralized decision-making, shaping governance structures within firms. Management literature has also emphasized that intermediate subsidiaries can serve as conduits for managerial coordination and knowledge flows within MNEs. Subsidiaries often act as critical nodes for internal knowledge transfer and coordination \citep{asmussen2011, gupta2000}. \cite{andersson2002} further highlight the strategic role of subsidiaries in developing capabilities and facilitating cross-unit learning within the group. \\


\noindent Indeed, governance concerns extend to the case of global ownership structures. \cite{aminadav2020corporate} examine patterns of corporate control around the world, focusing on ultimate controlling shareholders, and show that control is strongly correlated with shareholder protection, the stringency of labor contracts, and the power of unions. \cite{Fonseca_et_al_2023}  show that baseline gravity models are effective in predicting bilateral country linkages, with more populous, wealthy, and proximate countries more likely to have control relationships. They also highlight the role of shared legal traditions, cultural ties, and historical connections. Our paper builds on existing work by examining the internal mechanisms of corporate control, in particular how control is exercised within multinational firms through ownership chains, the role of intermediate affiliates, and the challenges posed by communication frictions and geographic distance. In practice, parent firms must decide how much control to exert over their foreign affiliates. They must balance internal monitoring with external contractual relationships. This decision is even more crucial when communication frictions and institutional weaknesses are considered. \cite{giroud2013proximity} demonstrate that headquarters tend to invest in subsidiaries that are geographically closer, due to lower monitoring costs, which in turn enhances productivity. \cite{chauvin2024working}  reveal that time-zone misalignment disrupts real-time coordination, increasing intra-firm communication costs. The role of contractual incompleteness is equally important. \cite{ottaviano2007distance} show that weak contract enforcement significantly influences firms' decisions on whether to own foreign affiliates or rely on arm’s-length relationships. When contracts are incomplete, MNEs tend to prefer direct ownership over outsourcing to mitigate hold-up risks, resulting in more hierarchical structures. This perspective extends previous governance theories by demonstrating that firms internalize operations not only to enhance knowledge flows but also to mitigate legal uncertainties and transactional risks.  Additionally, \cite{bloom2012organization} provide empirical evidence that firms with high internal trust are more likely to delegate decision-making, which affects hierarchical organization. \\

\noindent While corporations use hierarchical structures to maintain control and mitigate contractual risks, ownership chains are also shaped by regulatory considerations. The ability to shift profits across jurisdictions with different corporate tax rates provides a strong incentive for multinationals to create multi-layered ownership networks that facilitate the tax-efficient allocation of income. Unlike knowledge spillovers, which primarily improve internal efficiency, or control concerns, which focus on monitoring and governance, tax-motivated ownership complexity minimizes regulatory costs rather than operational coordination. \cite{grubert2003income}, highlighting the role of tax-motivated affiliate networks, shows that a substantial amount of income shifting by US MNEs occurs through intermediate holding companies that facilitate the routing of profits through conduit entities in tax-advantaged jurisdictions.  More recently, \cite{ferrari2024profit} develop a general equilibrium framework to analyze the impact of international tax rate differentials on corporate structuring. Their results show that the elasticity of shifted profits to tax rates is nearly three times larger than that of the tax base, underscoring how MNEs actively design ownership networks to minimize tax burdens rather than merely responding to production or operational needs. Empirical evidence confirms the large-scale impact of tax optimization on MNEs' structuring. \cite{laffitte2022profit} document how firms strategically shift both sales and profits to tax havens, often disproportionately reporting revenues in low-tax jurisdictions while real economic activity remains elsewhere.  Although tax optimization is a central driver of ownership complexity, other factors also play a role.\footnote{For example, \cite{lewellen2013internal} and \cite{dyreng2015effect} show that US multinationals often establish intermediate holding companies to facilitate regulatory arbitrage. See also \cite{francois2025tax} for recent evidence that complex ownership chains are associated with treaty shopping and tax-efficient investment routing.} Recent tax policy efforts, including the OECD's Global Minimum Tax initiative, aim to reduce tax incentives for complex ownership networks by limiting tax rate differentials across jurisdictions \citep{ferrari2024profit}. As global tax enforcement evolves, understanding the interplay between regulatory frameworks, tax planning, and deal structuring remains critical.\\

\noindent A key driver of multinational ownership is the need for hierarchical coordination of production and supply chains. Global companies structure their ownership networks to accommodate cross-jurisdictional interdependencies, ensuring operational efficiency while maintaining management control over dispersed operations. \cite{tintelnot2017global} develops a general equilibrium model showing that MNEs optimize production networks through trade-offs between the fixed costs of maintaining multiple locations and the efficiency gains from consolidating operations. Similarly, \cite{arkolakis2018innovation} and \cite{wang2021headquarters} demonstrate that MNEs make interdependent decisions regarding plant location, managerial control, and supply chain structure, which are driven by economies of scale and coordination efficiencies. To facilitate technology transfer, sourcing, and logistics, firms often rely on intermediary subsidiaries that help bridge geographic and institutional gaps. \cite{head2019brands} highlight how MNEs design supply chains to ensure brand consistency across markets, thereby reinforcing hierarchical control. These structures enhance continuity and efficiency in global value chains by reducing transaction costs associated with contractual incompleteness, supplier risk, and regulatory barriers.\footnote{A related framework by \cite{keller2013gravity} models firms' decisions to embed technological knowledge in traded intermediates or to transfer it directly to foreign affiliates.}\\

\noindent Against this background, our paper investigates the role of communication frictions in the formation and organization of multinational ownership networks.  We document the widespread diffusion of such structures and provide a simple theoretical framework, based on \cite{head2008fdi}, to rationalize the role of intermediary subsidiaries in managing and monitoring foreign affiliates in the presence of heterogeneous communication frictions. In doing so, we contribute to the growing literature on global firm organization and provide new insights into the determinants of multinational ownership structures. \\

\section{Mapping multinational ownership networks}\label{sec: data}

We construct a global dataset of multinational enterprises (MNEs) and their ownership structures using firm-level data from the Orbis database\footnote{The Orbis ownership database, a product of Moody's, has been widely used in research on MNEs. See, for example, \cite{Fonseca_et_al_2023}, \cite{Del_Prete_Rungi_2017}, and \cite{Cravino_Levchenko_2016}.}. Additional country-level covariates complement our analyses as needed, and they are described in Appendix Table \ref{tab: country covariates}. \\

\begin{figure}[H]
\centering
\caption{A fictional ownership chain}
\label{fig: fictional chain}
\includegraphics[width=.7\linewidth]{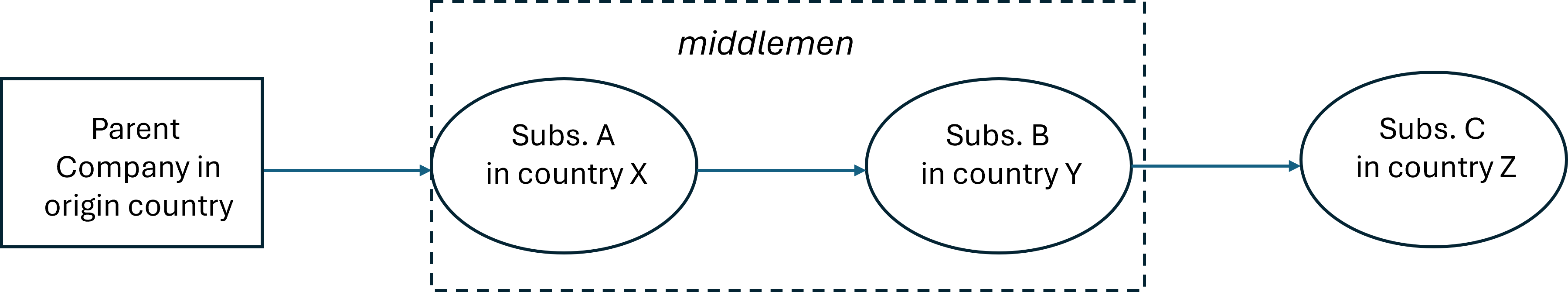}
\begin{tablenotes}
      \scriptsize 
      \item Note: Each node in the fictional graph represents a formally independent company, and each link is a stake that allows control of voting rights to the immediate upstream node. The parent company is a corporate shareholder on top of any ownership chain, while one or more \textit{middlemen} subsidiaries connect the parent company with a final subsidiary. Companies can be active in different countries.
      \end{tablenotes}
\end{figure}

\noindent When we refer to ownership structures, we mean the hierarchical control perimeter established by a parent company over its affiliates, which may be organized in ownership chains. The hypothetical chain of ownership in Figure \ref{fig: fictional chain} illustrates this concept. A subsidiary is directly controlled if the parent company holds an absolute majority of the voting rights in its shareholders' meeting (e.g. parent company directly controls subsidiary A in Figure \ref{fig: fictional chain}). Indirect control occurs when a subsidiary is owned through an intermediate affiliate that is itself controlled by the parent (e.g., Parent indirectly controls Subsidiaries B and C through an intermediary in Figure \ref{fig: fictional chain}).

\noindent To extract ownership structures from raw data, we apply the framework developed by \cite{rungi2017global}. This approach reconstructs ownership chains from the bottom up, starting from a global ownership matrix that includes all firms and their shareholders, without imposing a control threshold. Whenever a corporate shareholder appears in a firm's shareholder list, the algorithm initiates the detection of ownership chains.\footnote{In brief, the procedure identifies the parent company by tracing ownership structures from the bottom up, considering: (i) direct control, where a firm holds an absolute majority stake ($50\%$+1); (ii) indirect control by transitivity, where majority stakes align to form an ownership chain; and (iii) consolidated control, where fragmented stakes across subsidiaries collectively exceed the control threshold. Further details on the extraction of corporate control are provided in Appendix \ref{sec: appendix a}.  } For this study, we focus on multinational enterprises (MNEs), defined as parent firms that control at least one subsidiary located in a country other than their own.\\

\noindent Our final dataset includes 226,931 parent firms active in 2019 that collectively control 1,785,493 subsidiaries in 190 countries. Indirect control is pervasive among multinational enterprises: 54\% of foreign subsidiaries are controlled through at least one intermediate affiliate. At the same time, multinational groups characterized by complex ownership structures account for approximately 95\% of global sales by multinationals, underscoring their economic relevance\footnote{Appendix \ref{sec: appendix c} provides additional details on this evidence, as well as further descriptive statistics on the control networks of multinational enterprises in our sample.}.

\begin{table}[H]
\centering
\caption{Geographic distribution of MNEs by host economy in 2019: Parent companies, intermediate affiliates, and final subsidiaries}
\label{tab:geo cov mne}
\resizebox{.9\textwidth}{!}{%
\begin{tabular}{llrrlrrlrr}
\multicolumn{10}{c}{} \\ \hline
 &  & \multicolumn{1}{l}{} & \multicolumn{1}{l}{} &  & \multicolumn{1}{l}{} & \multicolumn{1}{l}{} &  & \multicolumn{1}{l}{} & \multicolumn{1}{l}{} \\
 &  & \multicolumn{1}{c}{\textbf{Parent Companies}} & \multicolumn{1}{c}{\textbf{\%}} &  & \multicolumn{1}{c}{\textbf{Total Subsidiaries}} & \multicolumn{1}{c}{\textbf{\%}} & \textit{of which} & \multicolumn{1}{c}{\textbf{Intermediate Affiliates}} & \multicolumn{1}{c}{\textbf{\%}} \\ \cline{3-4} \cline{6-7} \cline{9-10} 
 &  & \multicolumn{1}{l}{} & \multicolumn{1}{l}{} &  & \multicolumn{1}{l}{} & \multicolumn{1}{l}{} &  & \multicolumn{1}{l}{} & \multicolumn{1}{l}{} \\
\multicolumn{2}{l}{\textbf{EU27}} & \textbf{94,780} & \textbf{41.77\%} &  & \textbf{590,017} & \textbf{33.05\%} &  & \textbf{105,376} & \textbf{39.22\%} \\
 & \textit{of which} & \multicolumn{1}{l}{} & \multicolumn{1}{l}{} &  & \multicolumn{1}{l}{} & \multicolumn{1}{l}{} &  & \multicolumn{1}{l}{} & \multicolumn{1}{l}{} \\
 & Netherlands (the) & 11,060 & 4.87\% &  & 64,656 & 3.62\% &  & 17,215 & 6.41\% \\
 & Germany & 10,608 & 4.67\% &  & 107,659 & 6.03\% &  & 16,856 & 6.27\% \\
 & Italy & 7,729 & 3.41\% &  & 41,617 & 2.33\% &  & 7,192 & 2.68\% \\
 & France & 6,980 & 3.08\% &  & 51,774 & 2.90\% &  & 9,163 & 3.41\% \\
 & Spain & 5,654 & 2.49\% &  & 43,425 & 2.43\% &  & 6,928 & 2.58\% \\
 
 &  &  &  &  &  &  &  &  &  \\
\multicolumn{2}{l}{\textbf{Asia}} & \textbf{33,395} & \textbf{14.72\%} &  & \textbf{344,736} & \textbf{19.31\%} &  & \textbf{38,693} & \textbf{14.40\%} \\
 & \textit{of which} & \multicolumn{1}{l}{} & \multicolumn{1}{l}{} &  & \multicolumn{1}{l}{} & \multicolumn{1}{l}{} &  & \multicolumn{1}{l}{} & \multicolumn{1}{l}{} \\
 & China & 5,915 & 2.61\% &  & 127,154 & 7.12\% &  & 14,583 & 5.43\% \\
   & Japan & 5,109 & 2.25\% &  & 33,360 & 1.87\% &  & 2,241 & 0.83\% \\
& Singapore & 1,498 & 0.66\% &  & 35,547 & 1.99\% &  & 6,416 & 2.39\% \\
 & South Korea & 1,341 & 0.59\% &  & 5,353 & 0.30\% &  & 323 & 0.12\% \\
 &  & \multicolumn{1}{l}{} & \multicolumn{1}{l}{} &  & \multicolumn{1}{l}{} & \multicolumn{1}{l}{} &  & \multicolumn{1}{l}{} & \multicolumn{1}{l}{} \\
\multicolumn{2}{l}{\textbf{Other Europe}} & \textbf{30,495} & \textbf{13.44\%} &  & \textbf{204,749} & \textbf{11.47\%} &  & \textbf{40,188} & \textbf{14.96\%} \\
 & \textit{of which} & \multicolumn{1}{l}{} & \multicolumn{1}{l}{} &  & \multicolumn{1}{l}{} & \multicolumn{1}{l}{} &  & \multicolumn{1}{l}{} & \multicolumn{1}{l}{} \\
 & UK & 14,856 & 6.55\% &  & 133,366 & 7.47\% &  & 30,363 & 11.30\% \\
 & Switzerland & 8,790 & 3.87\% &  & 13,047 & 0.73\% &  & 2,873 & 1.07\% \\
  & Norway & 2,191 & 0.97\% &  & 17,273 & 0.97\% &  & 3,308 & 1.23\% \\
 &  & \multicolumn{1}{l}{} & \multicolumn{1}{l}{} &  & \multicolumn{1}{l}{} & \multicolumn{1}{l}{} &  & \multicolumn{1}{l}{} & \multicolumn{1}{l}{} \\
\multicolumn{2}{l}{\textbf{USA}} & \textbf{24,507} & \textbf{10.80\%} &  & \textbf{389,691} & \textbf{21.83\%} &  & \textbf{51,520} & \textbf{19.18\%} \\
 &  &  &  &  &  &  &  &  &  \\
\multicolumn{2}{l}{\textbf{Latin America}} & \textbf{22,805} & \textbf{10.05\%} &  & \textbf{80,875} & \textbf{4.53\%} &  & \textbf{8,015} & \textbf{2.98\%} \\
& \textit{of which} & \multicolumn{1}{l}{} & \multicolumn{1}{l}{} &  & \multicolumn{1}{l}{} & \multicolumn{1}{l}{} &  & \multicolumn{1}{l}{} & \multicolumn{1}{l}{} \\
 & The Caribbean & 17,831 & 7.86\% &  & 29,814 & 1.67\% &  & 4,552 & 1.70\% \\
& Brazil & 447 & 0.20\% &  & 11,786 & 0.66\% &  & 793 & 0.30\% \\  
   & Mexico & 312 & 0.14\% &  & 11,895 & 0.67\% &  & 583 & 0.22\% \\
 & Argentina & 134 & 0.06\% &  & 3,448 & 0.19\% &  & 197 & 0.07\% \\
 &  &  &  &  &  &  &  &  &  \\
\multicolumn{2}{l}{\textbf{Africa}} & \textbf{6,100} & \textbf{2.69\%} &  & \textbf{43,656} & \textbf{2.45\%} &  & \textbf{4,022} & \textbf{1.50\%} \\
 & \textit{of which} & \multicolumn{1}{l}{} & \multicolumn{1}{l}{} &  & \multicolumn{1}{l}{} & \multicolumn{1}{l}{} &  & \multicolumn{1}{l}{} & \multicolumn{1}{l}{} \\
   & South Africa & 725 & 0.32\% &  & 16,297 & 0.91\% &  & 2,351 & 0.88\% \\   
 & Nigeria & 115 & 0.05\% &  & 1,374 & 0.08\% &  & 60 & 0.02\% \\
 & Egypt & 154 & 0.07\% &  & 2,381 & 0.13\% &  & 114 & 0.04\% \\
 &  &  &  &  &  &  &  &  &  \\
\multicolumn{2}{l}{\textbf{Oceania}} & \textbf{6,034} & \textbf{2.66\%} &  & \textbf{57,071} & \textbf{3.20\%} &  & \textbf{11,589} & \textbf{4.31\%} \\
 & \textit{of which} & \multicolumn{1}{l}{} & \multicolumn{1}{l}{} &  & \multicolumn{1}{l}{} & \multicolumn{1}{l}{} &  & \multicolumn{1}{l}{} & \multicolumn{1}{l}{} \\
 & Australia & 4,336 & 1.91\% &  & 41,371 & 2.32\% &  & 9,418 & 3.51\% \\
 &  &  &  &  &  &  &  &  &  \\
\multicolumn{2}{l}{\textbf{Canada}} & \textbf{4,587} & \textbf{2.02\%} &  & \textbf{33,725} & \textbf{1.89\%} &  & \textbf{4,272} & \textbf{1.59\%} \\
 &  &  &  &  &  &  &  &  &  \\
\multicolumn{2}{l}{\textbf{Russia}} & \textbf{3,178} & \textbf{1.40\%} &  & \textbf{37,313} & \textbf{2.09\%} &  & \textbf{4,170} & \textbf{1.55\%} \\
 &  &  &  &  &  &  &  &  &  \\
\multicolumn{2}{l}{\textbf{Rest of the World}} & \textbf{1,050} & \textbf{0.46\%} &  & \textbf{3,660} & \textbf{0.20\%} &  & \textbf{804} & \textbf{0.30\%} \\
&  &  &  &  &  &  &  &  &  \\ \cline{3-4} \cline{6-7} \cline{9-10} 
\textbf{Total} & \textbf{} & \textbf{226,931} & \textbf{100.00\%} &  & \textbf{1,785,493} & \textbf{100.00\%} &  & \textbf{268,649} & \textbf{100.00\%} \\
 &  &  &  &  &  &  &  &  &  \\ \hline \hline
\end{tabular}%
}
\begin{tablenotes}
      \scriptsize 
      \singlespacing
      \item Note: The table reports the number of parent companies, intermediate affiliates, and final subsidiaries across different regions. Intermediate affiliates are subsidiaries that both control other entities and are themselves controlled within an ownership chain. The last columns highlight the number and share of intermediate affiliates, which serve as linking entities between parent companies and final subsidiaries. 
\end{tablenotes}
\end{table}

\noindent Table \ref{tab:geo cov mne} provides a geographic breakdown of parent companies and subsidiaries by hosting economy. The European Union (EU) accounts for 42\% of parent enterprises and 33\% of subsidiaries, reflecting the region's high degree of economic integration and geographical proximity. Within the EU, the Netherlands, Germany, Italy, France, and Spain host the largest number of both parent enterprises and subsidiaries. Asia is the second most important region with 14.7\% of parent enterprises and 19.3\% of subsidiaries. Among the Asian economies, China and Japan account for the largest shares of MNEs in our data set.

\noindent The United States has fewer parent firms than the European Union and Asia, but these firms control a larger number of firms. In particular, when we combine the United States and Europe, they account for about two-thirds of both parent firms and subsidiaries in our dataset. This pattern is consistent with aggregate FDI statistics, which show that developed economies attract the majority of global FDI inflows \citep{unctad2016inv}. \\

\noindent Turning to intermediate affiliates, shown in the rightmost columns of the table \ref{tab:geo cov mne}, we find that they are present in all economies, with no country hosting exclusively directly controlled affiliates. Generally, their distribution closely follows that of their parent enterprises. Two notable exceptions emerge, however. The United States and Australia host subsidiaries that are typically positioned within longer ownership chains, as indicated by their high ratios of intermediate affiliates to parent firms (2.10 and 2.17, respectively). This suggests that MNEs based in these countries rely more heavily on third countries and intermediate affiliates to structure their global operations, probably due to their geographical remoteness.  \\

The reliance on longer chains in the United States and Australia is consistent with the finding that geographic distance is associated with the use of intermediate affiliates, a pattern also observed in the subsequent empirical analysis. The following sections present stylized facts along these lines, preceded by brief remarks on tax optimization.

\subsection{Whither tax optimization?} \label{tax}

In this work, we investigate why one or more middlemen exist between a parent company and its subsidiaries. Such ownership configurations are commonly associated with tax optimization strategies and, thus, profit shifting. Specifically, refer to the work by \cite{francois2025tax} using the same data. They find that the number of layers of ownership between subsidiaries and the parent company correlates with a higher propensity to report zero profits. Moreover, MNEs with sufficiently complex ownership structures are more likely to shift profits away from their high-tax affiliates.\\


\noindent In the following paragraphs, we have a different perspective. We do not challenge the claim that profit shifting is more likely if ownership structures are complex. Yet, we believe that tax avoidance is not the main reason for the origin of such complex ownership structures. Indeed, if the locations of intermediary companies along an ownership chain were primarily driven by tax minimization, one would expect them to differ systematically from those of final subsidiaries, for instance, by being located in low-tax jurisdictions and exhibiting few signs of real economic activity. We estimate the likelihood that a middleman, compared to other subsidiaries, exhibits features typically associated with potential tax evasion practices. The traits we test include the strategic location in a tax haven and hallmarks of a Special Purpose Entity (SPE), such as the absence of employees and non-financial assets, established to obtain jurisdiction-specific advantages.

Column (1) of Table \ref{tab:th spe} reports predictions at the margin after a logistic regression in which the dependent variable is a dummy for tax-haven location, based on the classification of \cite{hines2010treasure}. The main explanatory variable is a dummy indicating middleman versus final subsidiaries. Firm age and size are included as controls, along with sector fixed effects.

\begin{table}[H]
\centering
\caption{Middlemen, tax havens, and SPEs}
\label{tab:th spe}
\resizebox{.8\textwidth}{!}{%
\begin{tabular}{lcc}
\hline \hline
& & \\
 & \multicolumn{2}{c}{Predicted probability of being:} \\[1ex]  \cline{2-3} 
 & (1) & (2) \\
 & \textbf{located in a tax haven} & \textbf{a suspected SPE} \\ \cline{2-3} 
 & \multicolumn{1}{l}{} & \multicolumn{1}{l}{} \\
Final subsidiary & 0.1445*** & 0.0042*** \\
 & (0.0016) & (0.0001) \\
 & \multicolumn{1}{l}{} & \multicolumn{1}{l}{} \\
Middleman subsidiary & 0.1628*** & 0.0054*** \\
 & (0.0017) & (0.0002) \\[1ex] \hline
 & \multicolumn{1}{l}{} & \multicolumn{1}{l}{} \\
Difference between predicted probabilities & 0.0183*** & 0.0012*** \\
 & (0.0015) & (0.0002) \\
 & \multicolumn{1}{l}{} & \multicolumn{1}{l}{} \\ \hline
\end{tabular}%
}
\begin{tablenotes}
 \footnotesize
     \item Note: The table reports predicted probabilities at the margin after two logistic regressions. In Column (1), the dependent variable is a dummy equal to one if the subsidiary is located in a tax haven; in Column (2), it is a dummy equal to one if the firm exhibits characteristics of a suspected special purpose entity. Predicted probabilities are obtain after estimates of coefficients for a dummy variable indicating whether the subsidiary is a middleman or a final subsidiary. All regressions control for firm size and age, and include sector fixed effects. Standard errors are clustered at the parent level (*** p$<$0.01, ** p$<$0.05, * p$<$0.1). 
\end{tablenotes}
\end{table}

Column (2) presents predictions at the margin from the same specification, but with a dummy indicating whether the subsidiary is suspected to be an SPE as the outcome. We follow the European Statistical System (ESS) definition of SPE\footnote{According to ESS, SPEs are legally established companies with up to five employees, little or no physical presence, and minimal production in the host economy. They are directly or indirectly controlled by foreign entities and transact almost exclusively with nonresidents.}, also referenced in the IMF guidelines \citep{Harutyunyan2022Special} and accordingly define potential SPEs as foreign subsidiaries with up to five employees, no non-financial assets, and no operating revenues. 

\noindent We find that the predicted probability for a middleman to be located in a tax haven is slightly higher than for a final subsidiary, 16\% vs 14\%. The difference between the two is statistically significant, albeit minimal\footnote{Please note that the location in a country that is classified as a tax haven is not \textit{per se} the indication that a firm engages in tax avoidance. }. Moreover, the likelihood of exhibiting features defining SPEs is very low for both types of subsidiaries. Altogether, results suggest that while tax optimization may partially explain the existence of multiple ownership chains, it accounts for only a small fraction of a broader, more complex reality. In other words, these patterns are not markedly more prevalent among middlemen than other subsidiaries, indicating that the use of middlemen for tax purposes is neither dominant nor exclusive.

\begin{figure}[H]
\centering
\caption{Corporate taxes along ownership chains}
\label{fig: tax variables}
\includegraphics[width=.6\linewidth]{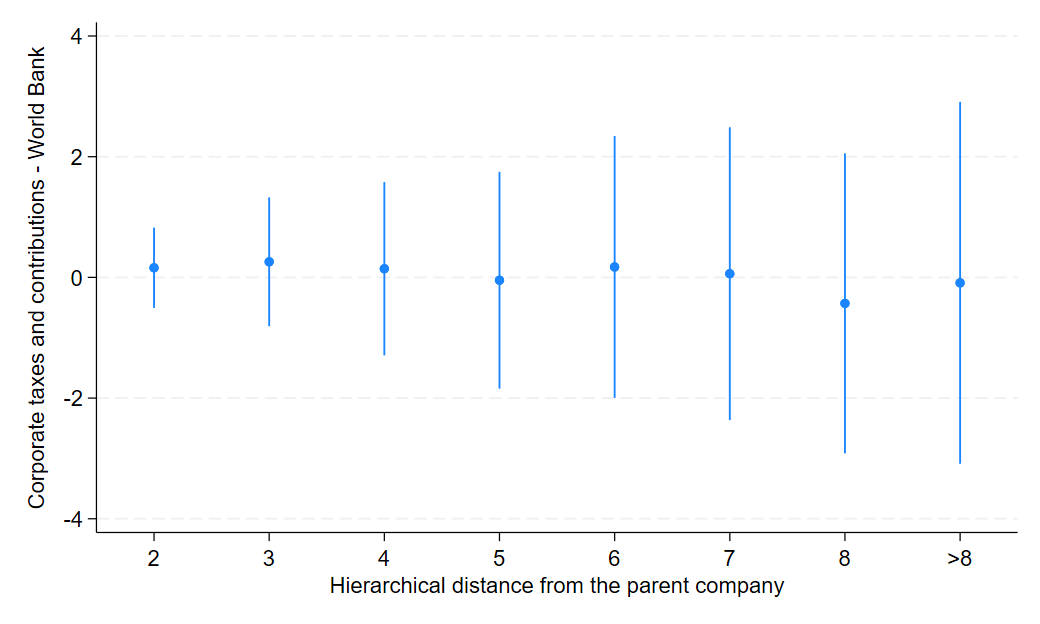}
\begin{tablenotes}
     \footnotesize
     \item Note: The figure presents the estimated coefficients of eq. \ref{eq:descriptive}, where the level of corporate income taxes in the country hosting subsidiaries is regressed on a categorical variable representing hierarchical distance from the parent company to the affiliate. Controls for the size and age of the subsidiaries are included, together with 2-digit industry fixed effects and parent-level fixed effects. The base category corresponds to hierarchical distance 1, where the parent directly controls the subsidiary. Consequently, all coefficients represent the variation in tax burden relative to this base category. Hierarchical distances above eight are grouped into a single category.
\end{tablenotes}
\end{figure}

\noindent We investigate this further by proposing a stylized fact that examines whether there is a correlation between the location along the ownership chain and lower corporate taxes in the hosting country. We propose a basic reduced form as follows:

 \begin{equation}
 \label{eq:descriptive}
   { Y_{p(i)s(j)}= \beta_{0} + \beta^{l} hier_{p(i)s(j)} + \beta_{1} X_{s(j)} + \epsilon_{p(i)s(j}}   
 \end{equation}

\noindent where $Y$ is the tax burden on a typical subsidiary firm $s$ in the hosting country $j$, which is controlled by a parent $p$ located in a country $i$. Country-level taxation is captured by the \cite{worldbank2020doingbusiness} total tax and contribution rate (TTCR). The main coefficient of interest, $\beta^{l}$, corresponds to a categorical variable measuring the hierarchical distance ($l$), i.e., the number of ownership links separating the subsidiary from the parent. Please note that we aggregate subsidiaries when they are located in layers above eight. The model also includes affiliate-level controls such as size and age in the matrix $X_{s(j)}$. Correlations are visualized in Figure \ref{fig: tax variables}. 

Evidently, we do not find any statistically significant pattern along the ownership chain. There is no apparent difference in tax burdens if a subsidiary is directly or indirectly controlled by the parent company. A typical intermediary subsidiary along the ownership chain would pay a similar amount of corporate taxes and mandatory contributions of any other subsidiary. 
\noindent Please note that we cannot confirm the absence of profit shifting or any other tax optimization strategy. What we can say is that the sequence along an ownership chain seems to have a different explanation. Yet, subsidiaries controlled by the same parent company might pursue tax optimization strategies regardless of their position along the ownership chain.    

\subsection{Ownership chains, distance, and overlapping working hours}

In this section, we highlight three key patterns that link ownership and geography. We begin by measuring the lengths of ownership chains.

\begin{itemize}
    \item \textsl{Ownership chains can be long, and they can cross multiple country borders. Approximately 61\% of subsidiaries are positioned at least two ownership links' distance from the parent company, 47\% cross at least two country borders, and 53\% extend approximately more than 10,000 kilometers.} 
\end{itemize}

\noindent A defining feature of the organizational complexity of multinational firms is their vertical structure, in which subsidiaries are arranged at multiple levels of the hierarchy, allowing the parent company to coordinate management decisions. 

\begin{table}[H]
    \centering
    \caption{How distant are subsidiaries from parent companies? Complex MNEs}
    \resizebox{.9\textwidth}{!}{
    \begin{tabular}{ccc|ccc|ccc}
    \hline
        Hierarchy & N. subs. & \%  & Countries & N. subs. & \%    & Distance (th. km) & N. subs. & \% \\ 
       (1) &  (2) & (3) & (4) & (5) & (6) & (7) & (8) & (9) \\  \hline
        1 & 350,228 & 39.00 & 1 & 475,491 & 52.95 & $\leq$ 5.0 & 413,546 & 47.29 \\ 
        2 & 247,358 & 27.55 & 2 & 335,588 & 37.37 & 5.0 - 7.5 & 202,758 & 23.18 \\ 
        3 & 133,535 & 14.87 & 3 & 67,892 & 7.56 & 7.5 - 10.0 & 105,782 & 12.1 \\ 
        4 & 73,380 & 8.17 & 4 & 15,461 & 1.72 & 10.0 - 12.5 & 66,684 & 7.62 \\ 
        5 & 40,893 & 4.55 & 5 & 2,683 & 0.3 & 12.5 - 15.0 & 37,441 & 4.28 \\ 
        6 & 21,949 & 2.44 & 6 & 687 & 0.08 & 15.0 - 17.5 & 39,891 & 4.56 \\ 
        7 & 13,204 & 1.47 & 7 & 158 & 0.02 & $>$17.5 & 8,458 & 0.97 \\ 
        8 & 7,395 & 0.82 &  &  &  &  &  &  \\ 
        9 & 4,292 & 0.48 &  &  &  &  &  &  \\ 
        10 & 2,545 & 0.28 &  &  &  &  &  &  \\ 
        11 & 1,377 & 0.15 &  &  &  &  &  &  \\ 
        12 & 908 & 0.10 & &  &  &  &  & \\ 
        $>$12 & 896 & 0.10 &  &  &  &  &  & \\ \hline
         &  &  &  &  &  &  &  &  \\ 
        domestic & 774,484 & 46.31 & domestic & 774,484 & 46.31 & domestic & 774,484 & 46.31 \\ 
      &  &  &  &  &  &  &  &  \\ \hline
        Total & 1,672,444 & 100.00 & Total & 1,672,444 & 100.00 & Total & 1,672,444 & 100.00 \\ \hline
    \end{tabular}
    }
    \label{tab: hier_km_dist}
    \begin{tablenotes}
      \scriptsize
      \singlespacing
      \item Note: The table reports descriptive evidence on foreign subsidiaries controlled by multinational enterprises. The simplest MNEs, which have only one direct subsidiary, are excluded. Columns 1, 2 and 3 on the left indicate the position in the ownership hierarchy, i.e. how many control links run from the HQs to the single subsidiary. Columns 4, 5, and 6, in the centre, count how many national borders are crossed along an ownership chain to reach a final subsidiary from the headquarters. Columns 7, 8, and 9 on the right report the geographic distance (thousands km) separating final subsidiaries from parent companies. Distance is measured between the most populated cities of hosting countries. 
\end{tablenotes}

\end{table}

\noindent To illustrate these patterns, we present the geography of the ownership chain in Table \ref{tab: hier_km_dist}, with the number of domestic affiliates listed separately at the bottom.  Columns 1 to 3 report the hierarchical distance of foreign affiliates from their parents, i.e., the number of ownership links required to reach a subsidiary from the headquarters.  While a relative majority (39\%) of foreign subsidiaries are directly controlled by the parent enterprise (hierarchical level 1, with no intermediate enterprises), the majority are located further down the hierarchy, at level 2 or above. However, as ownership chains lengthen, longer structures become increasingly rare, with only a small fraction of affiliates beyond the tenth hierarchical level.\footnote{The longest hierarchical distance observed in our dataset is 21 levels, indicating an ownership chain where 20 intermediate affiliates separate the headquarters from the final subsidiary.}

\noindent Beyond hierarchical structure, we also examine geographic distance along ownership chains, considering two complementary measures. Columns 4 to 6 capture the number of national borders crossed before reaching a subsidiary. Our findings show that most foreign subsidiaries (53\%) are located only one country away from their parent. Nevertheless, 37\% of the final subsidiaries are integrated into ownership chains that span at least two countries, while 7.6\% are located three countries away. Only in exceptional cases do parent firms establish ownership chains that extend up to seven countries (0.02\%) from the parent's home country.

\noindent Finally, we take a more classical approach by measuring the physical distance in kilometers between headquarters and subsidiaries, as reported in columns 7 to 9 of Table \ref{tab: hier_km_dist}. These data, obtained from \cite{CEPII} (CEPII database), represent the distance between the main cities in the countries where the headquarters and subsidiaries are located \footnote{Appendix Table \ref{tab: complex control chains} also shows the correlation between hierarchical distance of a subsidiary from the parent and the number of country borders crossed.}.

One could hardly expect that 47.3\% of the subsidiaries are located within 5,000 km of the parent company, a somewhat narrower range than expected. Beyond this threshold, the distribution has a long right tail, with subsidiaries located at increasing distances from the parent.\footnote{The longest observed distance in our dataset is 19,950 km, between Paraguay and Taiwan.} 

\begin{itemize}
    \item \textsl{Subsidiaries situated at more distant hierarchical levels in ownership chains are, on average, geographically farther from the parent company.}
\end{itemize}

\begin{figure}[H]
\centering
\caption{Geographical distance vs hierarchical distance of a parent and its affiliates}
\label{fig: km_hier_dist}
\includegraphics[width=.55\linewidth]{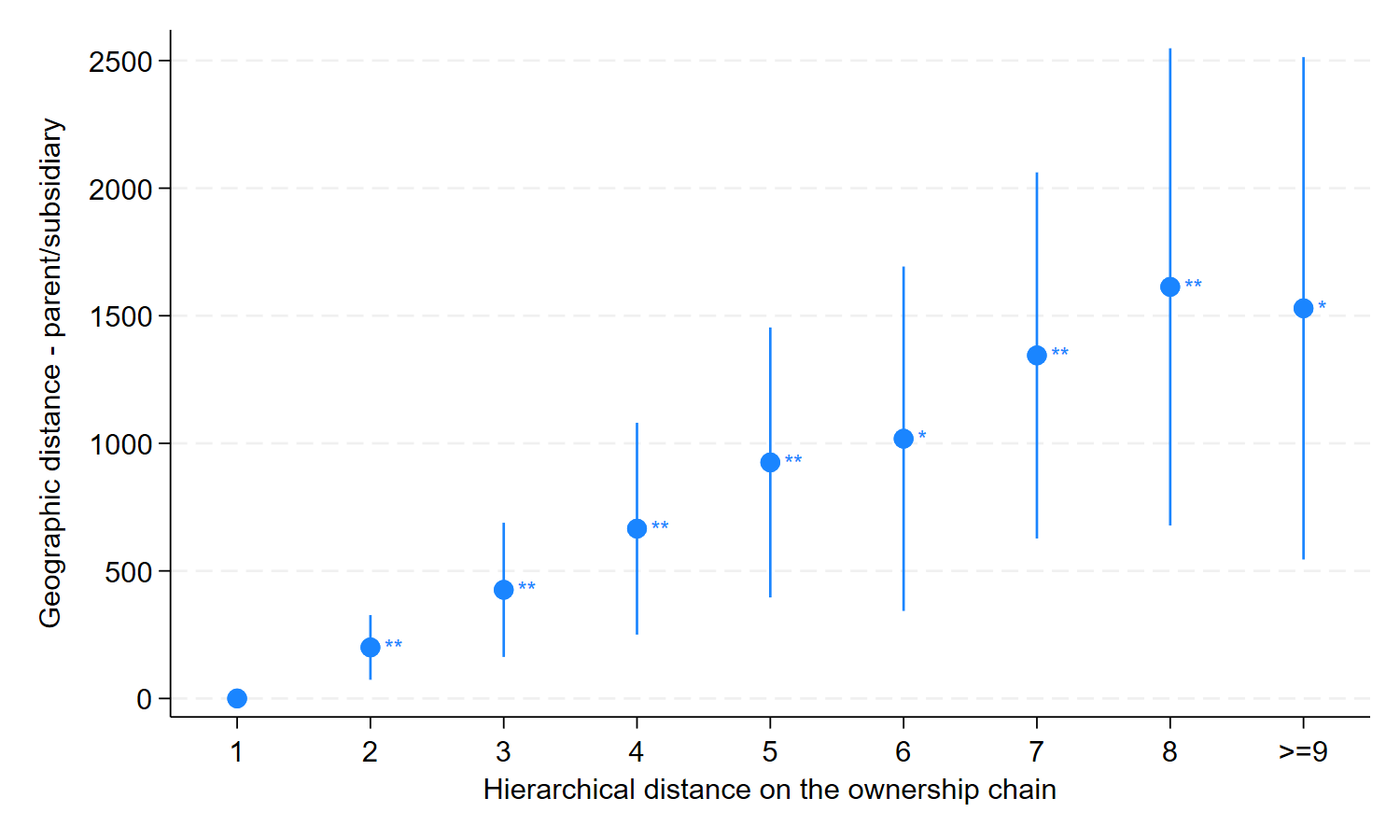}
\begin{tablenotes}
     \footnotesize
     \item Note: The figure presents the estimated coefficients after the reduced-form eq. \ref{eq:descriptive}, where the geographical distance between parent companies and their affiliates is regressed on a categorical variable representing hierarchical distance. Parent-company fixed effect is included, along with controls for the size and age of intermediary affiliates. The base category corresponds to hierarchical distance 1, where the parent directly controls the subsidiary. Consequently, all coefficients represent the variation in geographical distance relative to this base category. Hierarchical distances above eight are grouped into a single category. Standard errors, clustered at the parent-company level, are reported in parentheses (*** p$<$0.01, ** p$<$0.05, * p$<$0.1).
\end{tablenotes}
\end{figure}

\begin{figure}[H]
\centering
\caption{Overlapping working hours vs hierarchical distance of a parent and its affiliates}
\label{fig: wh_km_dist}
\includegraphics[width=.55\linewidth]{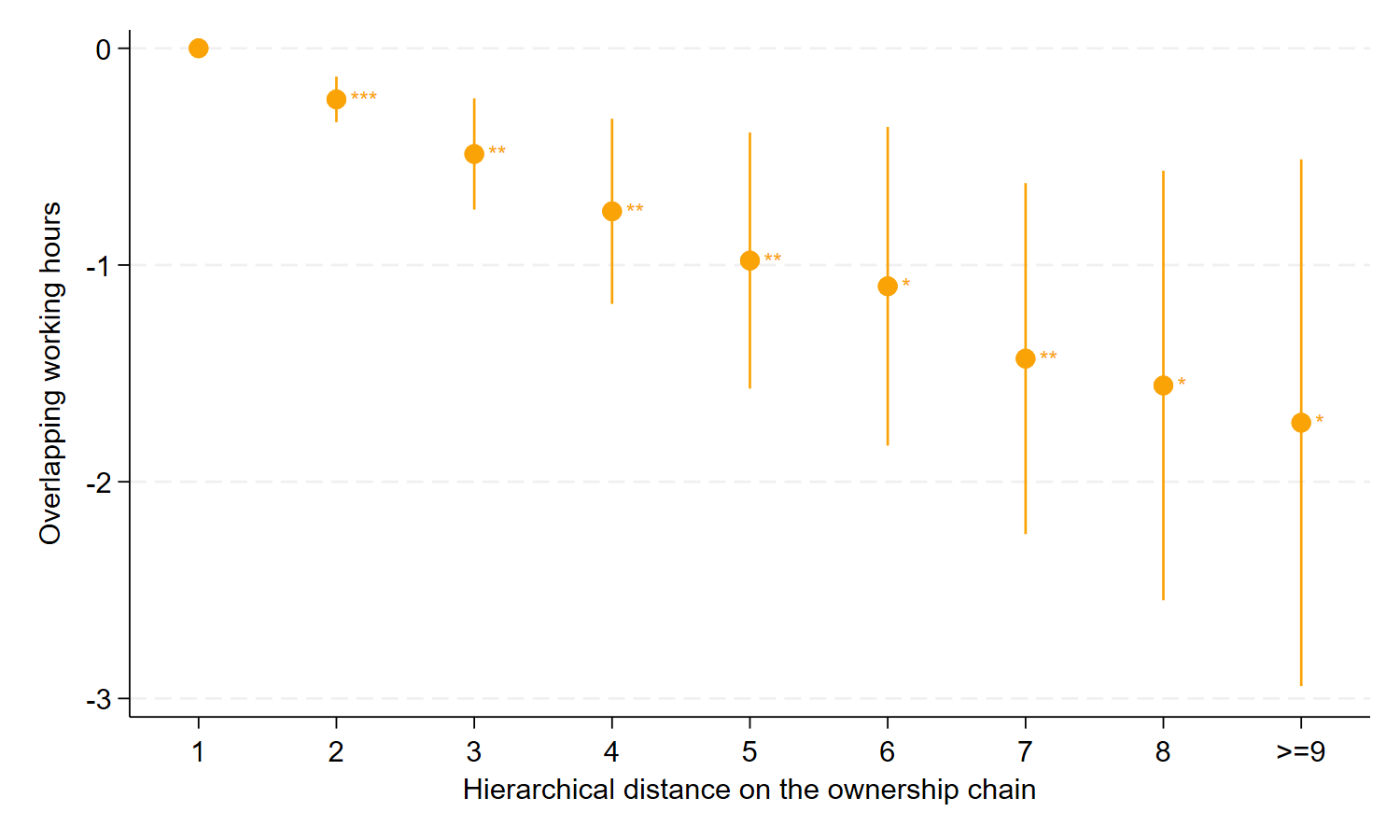}
\begin{tablenotes}
      \footnotesize 
      \item Note: The figure presents the estimated coefficients after the reduced-form eq. \ref{eq:descriptive}, where overlapping working hours—used as a proxy for communication frictions—are regressed on a categorical variable representing hierarchical distance between parent companies and their subsidiaries. A parent-company fixed effect is included, along with controls for subsidiary size and age. The base category corresponds to hierarchical distance 1, where the parent directly controls the subsidiary. Consequently, all coefficients represent the variation in overlapping working hours relative to this base category. Hierarchical distances above eight are grouped into a single category. Standard errors, clustered at the parent-company level, are reported in parentheses (*** p$<$0.01, ** p$<$0.05, * p$<$0.1). 
\end{tablenotes}
\end{figure}

\noindent We illustrate these patterns in Figures \ref{fig: km_hier_dist} and \ref{fig: wh_km_dist}, where we analyze the relationship between the affiliate's position in the ownership chain and its geographic distance from the parent. The relationship is estimated by means of the equation \ref{eq:descriptive} used in the previous paragraphs, where $Y$ represents either the geographical distance (in thousands of km) or the number of overlapping working hours between the parent in country $i$ and its affiliate in country $j$. The latter is defined as the total number of hours within a standard business day during which the working schedules of two firms’ locations overlap, given their respective local time zones and assuming a ten-hour working day. The main coefficient of interest, $\beta^{l}$, corresponds to a categorical variable measuring the hierarchical distance ($l$), i.e., the number of ownership links separating the subsidiary from the parent. The model also includes affiliate-level controls such as size and age ($X_{s(j)}$), while parent-level fixed effects ($\delta_{p(i)}$) allow us to capture variation within MNEs.

\noindent In Figure \ref{fig: km_hier_dist}, we present the estimated coefficients ($\beta^{l}$) along with their confidence intervals, illustrating the relationship between hierarchical position and geographic distance. We find a positively increasing and statistically significant distance premium relative to the base category (first-tier subsidiaries). In other words, subsidiaries positioned further down the hierarchy tend to be geographically further away from the parent company. However, the degree of variation is not negligible, as the confidence intervals widen at lower hierarchical levels where there are fewer and fewer affiliates.

\begin{itemize}
    \item \textsl{Subsidiaries situated at more distant hierarchical levels in ownership chains have, on average, less overlapping working hours with the ultimate parent company.}
\end{itemize}

\noindent Figure \ref{fig: wh_km_dist} reveals a clear hierarchical gradient in time overlap. As subsidiaries sit lower in the ownership chain, their overlapping working hours with the ultimate parent decline monotonically. This pattern carries sharp implications for coordination: subsidiaries at the tail of long chains are more likely to operate outside the parent’s business hours, making direct managerial oversight costly, and increasing the need for delegation to intermediate entities.\\

\noindent We provide further evidence consistent with this intuition by estimating a naive control gravity equation that isolates the role of overlapping working hours across firm locations. We aggregate ownership links at the country-pair level, distinguishing parent–subsidiary from middleman–subsidiary relationships. As illustrated in Figure \ref{fig: fictional chain}, links from the parent to both intermediaries A and B and final subsidiary C define parent control, while links from intermediaries A and B to the final subsidiary C capture middleman control. Table \ref{tab: bigravity} reports estimates from the following reduced-form equation:

 \begin{equation}
 \label{eq:bigravity}
   { N_{ij}= exp(\beta^{wh} wh_{ij} + \boldsymbol{\beta'}\mathbf{x}_{ij} +\gamma_i+\gamma_j) \epsilon_{ij}}   
 \end{equation}
where $i$ and $j$ stand for origin and destination country and $N_{ij}$ counts the number of companies in country $j$ controlled by companies located in country $i$. $wh_{ij}$ measures overlapping working hours between $i$ and $j$, while $\mathbf{x}_{ij}$ is a vector of gravity controls we source from the CEPII Gravity dataset (namely, the logarithm of distance measured in km, and indicator variables for the presence of, respectively, contiguous borders, common language, common legal origins, colonial ties and regional trade agreements between countries). $\gamma_i$ and $\gamma_j$ capture origin and destination country fixed effects.

We find that overlapping working hours are significant only in the control relationships between middlemen and final subsidiaries. If real-time communication is central to organization and coordination, this constraint does not bind the location of subsidiaries relative to the parent. Instead, coordination costs appear to be shifted downstream, with middlemen absorbing the burden of synchronizing operations on behalf of the parent.

\begin{table}[H]
\centering
\caption{Two naive control gravity equations for MNEs}
\label{tab: bigravity}
\resizebox{.8\textwidth}{!}{%
\begin{tabular}{lcc}
\hline
 &  &  \\
\textbf{Dep. Var.} & \multicolumn{2}{c}{$N_{ij}$= \# of companies country $i$ controls in country $j$} \\ \hline
 &  &  \\
 &  &  \\
Sample & \begin{tabular}[c]{@{}c@{}}Parent-subsidiary\\ control links\end{tabular} & \begin{tabular}[c]{@{}c@{}}Middleman-final\\ control links\end{tabular} \\ \hline
 &  &  \\
N. of overlapping & 0.006 & 0.068** \\
working hours & (0.023) & (0.027) \\
 &  &  \\
RTA & 0.136 & 0.291* \\
 & (0.195) & (0.168) \\
Log distance (km) & -0.336*** & -0.227*** \\
 & (0.072) & (0.088) \\
Home & 2.516*** & 3.472*** \\
 & (0.315) & (0.250) \\
Language & 0.747*** & 0.641*** \\
 & (0.089) & (0.108) \\
Colony dependence & 0.312** & 0.385*** \\
 & (0.129) & (0.124) \\
Legal origins & 0.238*** & 0.268*** \\
 & (0.087) & (0.101) \\
 &  &  \\ \hline
Observations & 32,952 & 35,012 \\
Fixed Effects & i,j & i,j \\ \hline
\end{tabular}%
}
 \begin{tablenotes}
      \scriptsize
      \singlespacing
     \item Notes: $N_{ij}$ counts the number of companies in country $j$ controlled by companies in country $i$. Results in Column (1) refer to the sample of parent-subsidiary control relationships, whereas Column (2) refers to middleman-final control relationships. Standard errors clustered by origin country in parentheses (*** p$<$0.01, ** p$<$0.05, * p$<$0.1).
 \end{tablenotes}
\end{table}

\noindent These findings are central to our structural model presented in Section \ref{sec: model}. The core intuition is that parent firms structure ownership chains to facilitate efficient managerial coordination. When subsidiaries are geographically distant, direct supervision becomes impractical, necessitating delegation. If this hypothesis holds, then the observed structure of ownership chains reflects the communication needs of managers at different hierarchical levels, providing a rationale for the relationships documented in Figures \ref{fig: km_hier_dist} and \ref{fig: wh_km_dist}.

\section{An analytical framework for ownership chains}
\label{sec: model}

In this section, we provide a conceptual framework to explain location strategies along ownership chains in multinational enterprises (MNEs). We extend the \cite{head2008fdi} model to incorporate the possibility of ownership chains. The original \cite{head2008fdi}' setup assumes that headquarters compete to win control over new subsidiary assets worldwide through an auction, while incurring monitoring costs. In the context of ownership chains, we posit that each headquarters faces a choice: it can either directly monitor a final target subsidiary or delegate monitoring to an intermediate managerial unit, an intermediate affiliate (sometimes referred to as a middleman). These intermediate affiliates are subsidiaries that are already part of the corporate structure prior to the auction taking place \footnote{In principle, intermediate affiliates could be monitored themselves. However, incorporating recursive monitoring would complicate the model without generating additional testable implications for the initial allocation of control along the ownership chain. Therefore, within the present framework, intermediate affiliates are best understood as organizational nodes to which monitoring authority is delegated, rather than as agents whose incentives are the primary focus of this study. More generally, delegation of monitoring authority is itself a meaningful organizational margin, even when monitoring technologies are identical, because it reshapes incentive and commitment constraints \citep{Strausz1997}.}. A subsidiary is added to an ownership chain when the headquarters delegates or self-delegates monitoring, i.e., when the cost of delegation is effectively zero. In the latter case, we have the special case of direct control.

We conceptualize the investment decision process as unfolding in two stages. First, the headquarters bids in an auction for a new target subsidiary, contingent upon the inspection costs given the target's location. Second, if it wins the auction, it delegates the unit located in an intermediate jurisdiction that minimizes inspection costs. As a result, inspection costs affect the bidding strategy in the auction. Importantly, headquarters always retains the option to monitor subsidiaries directly, in which case no intermediate affiliate is necessary.

 Note that the model underlying our analysis abstracts from production processes between headquarters and subsidiaries. While our setup does not exclude the possibility of supply chain coordination, it also applies to conglomerates and diversified business portfolios, where monitoring remains crucial for governance and managerial control.\footnote{ To assess whether the length of ownership chains is associated with increasing distance in the production space, we ran a regression in which the hierarchical distance of an affiliate from its parent served as the dependent variable, including, among other controls, the ``production distance'', i.e. $upstreamness_{ij}$, a measure from \cite{antras2019}. While the coefficient was statistically significant, its magnitude was negligible. This suggests that an implausibly long ownership chain would be required for an economically meaningful difference in production stages between parent and subsidiary. In our data, these results suggest that ownership chains are primarily shaped by organizational considerations rather than sequential production optimization. Results are available upon request.}\\

\cite{head2008fdi} formalize the decision-making process introducing a classical inspection game between the headquarters (Hq) and a target subsidiary (Sub). This framework helps headquarters determine the maximum bid they can offer in auctions for new subsidiaries. The possible strategies and corresponding payoffs for both players are summarized in Table \ref{tab: game}.\footnote{A textbook version of this inspection game was originally introduced by \cite{Fudenberg_Tirole1991} and later adapted to the FDI context by \cite{head2008fdi}. We adapt their framework by incorporating the possibility of an intermediate monitoring unit within the multinational firm’s ownership network.}

\noindent In this corporate governance game, profits arise from the value generated by the headquarters (Hq) and its existing perimeter of subsidiaries ($A$)\footnote{The existence of a corporate perimeter before new acquisitions implies that $A= \sum_f a_f$, meaning that the total value is generated by both Hq and all existing subsidiaries, regardless of new acquisitions.}, as well as by the additional subsidiary ($b$), conditional on the effort exerted by the subsidiary's manager ($e$). Headquarters can choose to monitor the subsidiary with probability $q$ or trust it with probability $1-q$ . Similarly, the subsidiary's manager can work ($1-p$) or shirk ($p$). If the subsidiary works, headquarters pays a wage $w$; however, if headquarters monitors and discovers shirking, the wage and the value of the subsidiary default to zero ($w = b = 0$).

       \begin{table}[H]
       \caption{The inspection game}
       \label{tab: game}
     \centering
    \begin{tabular}{ccc|c|c|}
      & & \multicolumn{1}{c}{} & \multicolumn{2}{c}{\textbf{Sub} }\\
      & & \multicolumn{1}{c}{} & \multicolumn{1}{c}{$Work$}  & \multicolumn{1}{c}{$Shirk$} \\
            & & \multicolumn{1}{c}{} & \multicolumn{1}{c}{$(1-p)$}  & \multicolumn{1}{c}{$p$} \\\cline{4-5}
      \multirow{2}*{\textbf{Hq} }  & $Trust$ &(1-q)& $A+b-w, w-e$ & $A-w ,w$ \\\cline{4-5}
      & $Monitor$ & ($q$) & $A+b-w-C,w-e$ & $A-C,0$ \\\cline{4-5}
    \end{tabular}
  \end{table}  

\noindent The headquarters (Hq) bears an inspection cost ($C$) to verify the activities of the subsidiary (Sub), a concept that we elaborate on in the next section. In the classical version of this game, we assume $b \geq w \geq e \geq C \geq 0$, which results in no pure-strategy Nash equilibrium. Therefore, in a mixed-strategy equilibrium, the authors obtain a final valuation function that Hq assigns to Sub:

\begin{equation}
    \label{eq:vf}
v=A+ b -2\sqrt{bC}    
\end{equation}

Since corporate control operates in a global market, all Hqs play the same inspection game when evaluating subsidiaries that are available for acquisition. Consequently, competition ensures that the winning Hq is the one offering the highest valuation, determined by equation \ref{eq:vf}. 


With respect to the original model, we assume that, when acquiring a subsidiary (Sub), headquarters (Hq) can choose between two monitoring strategies: (i) direct monitoring, where Hq monitors the subsidiary itself; or (ii) delegated monitoring, where Hq delegates the monitoring task to an intermediate affiliate (Mid). There are costs associated with both monitoring and delegation. More specifically, a Hq in country $i$ that wants to monitor a Sub in country $j$ wants to minimize the following composite cost function:

\begin{equation}\label{eq: composite cost}
    C_{ikj}=\delta_{ik}+\delta_{kj}-\epsilon_{ikj}
\end{equation}

\noindent where: $\delta_{kj}$ represents the cost incurred by a Mid in country $k$ to monitor a Sub in country $j$; $\delta_{ik}$ represents the cost incurred by Hq in country $i$ to delegate monitoring to a Mid in country $k$;  $\epsilon_{ikj}$ is a stochastic unobservable factor specific to the $ikj$-th ownership chain that captures idiosyncratic variations in the business environment that may increase or decrease monitoring costs. Notably, we assume, realistically, that delegated monitoring is always less expensive than direct monitoring, such that: $\delta_{kj} \geq  \delta_{ik} $>$ 0$. 

In addition, the inspection game also applies to domestic targets. Each headquarters competes in a global auction for all available subsidiaries, including those in its home country. Thus, this framework accounts for ownership chains composed of both foreign and domestic subsidiaries. \\

\noindent Domestic subsidiaries require a separate discussion because they introduce three different location strategies:

\begin{enumerate}
    \item Headquarters (Hq) and an intermediate affiliate (Mid) are both located in the same home country ($i$) and control a foreign subsidiary (Sub) in country ($j$). This scenario creates a domestic segment within the ownership chain, where the cost structure follows: $\delta_{kj} \geq (\delta_{ik}=\delta_{ii}) > 0$. This implies that delegating monitoring to a domestic Mid remains cheaper than direct monitoring by Hq, even within the same country.
    \item All three entities — Hq, Mid, and Sub — are located within the same home country ($i$). In this case, the composite cost function simplifies to: $2\delta_{ii} > 0$. Here, monitoring and delegation costs are identical but remain greater than zero because domestic frictions -- although smaller than foreign frictions -- are still present. Since internal distances differ between countries, geography alone does not determine the chain structure. In the absence of other cost factors, Mid and Sub could theoretically swap positions within the ownership chain without affecting total costs, meaning that only the sequence of acquisitions determines their relative placement.
    \item In rare cases, the Hq in country ($i$) delegates monitoring to a foreign Mid in country ($k$), which then monitors a domestic Sub in ($i$). This results in a roundtripping strategy, where an offshore intermediate affiliate is placed in the chain even though the final subsidiary is located in the same country as the parent. One possible explanation is that domestic investors use offshore intermediaries to shelter their investments from the public authorities through foreign shell companies. However, our model did not account for these cases.\footnote{According to our data, we observe 21,624 cases of roundtripping, where at least one intermediary is located abroad, but the final subsidiary remains in the same country as the parent. These cases represent about 0.01\% of all final subsidiaries.}
\end{enumerate}

Finally, our framework also accounts for directly controlled subsidiaries. In fact, the simplest strategy for Hq is to retain full corporate control and not delegate monitoring to a Mid. In this case, Hq self-delegates, which means that delegation costs are effectively zero. More formally, in our model, direct control is treated as a valid location choice without delegation, represented as a $ikj$-th triplet where the intermediate jurisdiction ($k$) is null, i.e., $k = \emptyset$.

\subsection{A two-step investment decision}

Building on the framework introduced in the previous sections, we derive a system of two equations that model the two-step investment decision process: (i) a triangular equation that determines the location of an intermediate affiliate (Mid), conditional on the location of final subsidiaries (Subs); and (ii) a bilateral equation that determines the location of final subsidiaries (Subs).

\subsubsection{Locations of the chain}

We assume that $\epsilon_{ikj}$ is independently and identically distributed (i.i.d.) following a Type I extreme value distribution. The probability that Hq in country $i$ selects country $k$ as the monitoring location, conditional on investing in country $j$, corresponds to the probability that $C_{ikj}$ is minimized. This yields:

\begin{equation*}
     \pi_{ik \mid j} = \mathds{P}(\mathcal{M}_{ik}=1)  =\mathds{P}(C_{ikj}\leq C_{i\ell j}, \forall \ell \neq k)  
\end{equation*}

After some algebraic derivations, we obtain the following result\footnote{All intermediate steps are available upon request.}

\begin{equation*}
      \mathds{P}(\mathcal{M}_{ikj}=1) = \frac{e^{-(\delta_{ik}+\delta_{kj})}}{\sum\limits_{\ell} e^{-(\delta_{i\ell}+\delta_{\ell j})}} \\
\end{equation*}

We now derive the first testable equation, which characterizes the location choice for ownership chains.\\

\textbf{I. Locations of ownership chains} \textit{The probability that Hqs in country $i$ select country $k$ as the monitoring location for a Sub in country $j$ is given by:} 

\begin{equation}
\label{eq:prcond}
    \begin{aligned}
    \pi_{ik \mid j} = \mathds{P}(\mathcal{M}_{ikj}=1) = \frac{e^{-(\delta_{ik}+\delta_{kj})}}{\sum\limits_{\ell} e^{-(\delta_{i\ell}+\delta_{\ell j})}}
    \end{aligned}
\end{equation}\\
 
\noindent where we can define $\mathcal{C}_{ij} = -ln [ \sum\limits_{\ell} e^{-(\delta_{i\ell}+\delta_{\ell j})}]$ as the expected aggregate cost of locations $j$ for Hqs in country $i$, whatever the intermediate country $k$.

\subsubsection{Probability of winning the auction for corporate control}

Headquarters (Hqs) compete for control of new subsidiaries (Subs) by participating in an auction. Each Hq places a bid for each Sub based on its valuation function: $v=A + b -2\sqrt{b C}$.


The highest valuation wins the auction, i.e. the marginal probability of a Hq in country $i$ gaining control over a Sub in country $j$ is given by the probability that the highest valuation for a given Sub in $j$ comes from an Hq located in $i$, i.e. the maximum valuation in $i$ is the highest among the maxima of all competing countries. This probability is:

\begin{equation*}
    \pi_{ij} = \mathds{P}(\mathcal{N}_{ij}=1) =
\end{equation*}

\begin{equation*}
           = \mathds{P}(a_n^{max} \leq a_i^{max} -2\sqrt{b C_{ij}} + 2\sqrt{b C_{nj}}, \forall n \neq i)
\end{equation*}

\noindent where $\mathcal{N}_{ij}=1$ if an Hq in country $i$ wins control of a Sub in country $j$. We denote by $m_i$ the number of Hqs in country $i$. After some algebraic derivations, we derive the second testable equation\footnote{All intermediate steps are available upon request.}.

\textbf{II. Probability of controlling subsidiaries} \textit{The probability that a Hq located in country $i$ wins the auction for the control of a Sub located in $j$ is given by:}

\begin{equation}
\label{eq:prmarg}
    \begin{aligned}
       \pi_{ij} &= \frac{m_i \ \exp \biggr(  -(2\sqrt{b C_{ij}})/\sigma + \mu_{i}/\sigma \biggr)}{\sum\limits_n m_n \exp \biggr( -(2\sqrt{b C_{nj} })/\sigma + \mu_{n}/\sigma \biggr) }
    \end{aligned}
\end{equation}\\

\section{Empirical strategy}
\label{sec: str}

Equations \ref{eq:prcond} and \ref{eq:prmarg} constitute the foundation of our empirical strategy. The first equation describes the parent's strategy for locating new subsidiaries along ownership chains, while the second characterizes the probability that the parent will gain control of a subsidiary in a given location.

\noindent We assume that $\pi_{ik \mid j}$ is identical across parents, which allows us to derive an aggregate form of the testable equation. By summing individual location choices, we define $M_{ikj}$, which represents the number of parents in country $i$ that delegate monitoring responsibilities to an intermediate affiliate in country $k$ for final subsidiaries located in country $j$.\footnote{Note that direct control is a special case of an ownership chain in which the parent self-delegates monitoring. In this case, $k$ represents a null country and the delegation cost is zero.} We obtain the expected value of $M_{ikj}$ as:

    \begin{equation*}
    \label{eq:indcount}
    \begin{aligned}
            \mathds{E}[M_{ikj}] &= \pi_{ik \mid j} M_{ij} = exp(-\delta_{ik}-\delta_{kj}+\mathcal{C}_{ij}+\ln M_{ij})
    \end{aligned}
\end{equation*}

From the above empirical equation, we derive this triangular gravity equation: 

    \begin{equation}
    \label{eq:triangular structural}
    \begin{aligned}
           \mathds{E} \biggr[\frac{M_{ikj}}{M_{ij}}\biggr] = exp( \alpha_1 X_{ik} + \alpha_2 X_{kj} + FE_{ij})
    \end{aligned}
\end{equation} 

where the result represents the share of triplets ($ijk$) among all observed triplets, and $\beta$ and $\rho$ represent our coefficients of interest. $X_{ik}$ and $X_{kj}$ are vectors of geographic and institutional control variables that capture factors potentially hindering coordination along ownership chains.

Beyond standard variables, such as bilateral distance, contiguity, common language, shared legal origin, colonial ties, and participation in regional trade agreements, these vectors also embed a key measure of organizational frictions: the number of overlapping working hours.

Building on \cite{stein2007longitude} and \cite{bahar2020hardships}, we interpret time-zone misalignment as a salient barrier to real-time communication. Limited overlap in working hours between the parent country $i$ and the intermediate affiliate $k$, as well as between $k$ and the final subsidiary $j$, raises coordination costs and can significantly shape the structure of multinational ownership chains. Finally, we include the bilateral fixed effect ($FE_{ij}$), which captures the expected aggregate cost index, $\mathcal{C}_{ij}$, as from the denominator of Equation \ref{eq:prcond}. 

\noindent Having established the triplet structure, we turn to the final investment decision, described in equation \ref{eq:prmarg}. We assume that the probability of choosing destination $j$ is constant across parent firms in a given country $i$. Investment decisions are then aggregated into the variable $M_{ij}$, which represents the total number of new subsidiaries in $j$ held by parents in $i$, either through direct or indirect control paths. The expression for the expected value of $M_{ij}$ is given by:

\begin{equation*}
    \begin{aligned}
             \mathds{E}[M_{ij}] &= \pi_{ij} M_{j} \\    
            &= exp \biggr(  \ln \Bigr(\frac{m_i}{\sum_n m_n} \Bigr) + \frac{\mu_i}{\sigma} + \ln M_j - \ln S_j -(2\sqrt{b E[C_{ikj}]})/\sigma \biggr)
    \end{aligned}
\end{equation*}

\noindent where $S_j$ is the weighted average productivity level of competing firms from other countries, adjusted for the cost they incur in overseeing activities in $j$. This makes $S_j$ a measure of the degree of competition to acquire assets in market $j$. Meanwhile, $M_j$ is the total number of subsidiaries operating in $j$, and $E[C_{ikj}]$ is the expected value of the inspection cost of a Hq in country $i$ that controls a Sub in country $j$, having chosen a representative $k$.

\noindent To control for country-specific factors, we include fixed effects for source and destination countries in equation \ref{eq:prmarg}. The home country fixed effect ($FE_i$) captures the share of parent companies headquartered in $i$, along with their average productivity. The destination market fixed effect ($FE_j$) reflects the size and competitiveness of the market, including both the number of affiliates and the degree of competition from other firms.

\noindent After substituting $E[C_{ikj}]=\frac{\hat{\mathcal{C}}_{ij}}{K}$, where $\hat{\mathcal{C}}_{ij}$ is the estimate of the bilateral fixed effect from the first step, and imposing $\theta=\frac{2\sqrt{b}}{\sigma}$, we obtain the structural equation for final investments:


    \begin{equation}
    \label{eq:bilateral structural}
    \begin{aligned}
           \mathds{E}[M_{ij}^A] = exp(- \theta \sqrt{ \frac{\hat{\mathcal{C}}_{ij}}{K}} +  FE_i + FE_j)
    \end{aligned}
\end{equation}

\noindent where $\theta$ is our coefficient of interest, and $FE_{i}$ and $FE_{j}$ represent origin and destination fixed effects, respectively. Finally, equations \ref{eq:triangular structural} and \ref{eq:bilateral structural} form a system of two testable equations, which we estimate using a Pseudo-Poisson Maximum Likelihood (PPML) estimator, following \cite{silva2006log}.

\section{Results}
\label{sec: res}

Table \ref{tab:structural gravity} reports estimates for equations \ref{eq:triangular structural} and \ref{eq:bilateral structural} across three nested specifications, moving from conventional gravity predictors to variables more closely related to cross-border communication and coordination.

Columns (1), (3) and (5) show results for the triangular equation specified in equation \ref{eq:triangular structural}. In the first column we present a restricted specification that includes geographical distance (measured in kilometers), a dummy for the presence of a regional trade agreement (RTA), and a dummy indicating domestic control relationships.  We see that the farther the middleman is from either the parent or the final destination, the smaller the share of investments routed through $k$, with the effect being stronger for distance to the final destination ($-26\%$) than for distance to the parent ($-5.6\%$). Both the RTA and the domestic-control dummies have positive and statistically significant effects.\\
 
Column (3) augments the model with standard gravity variables that capture information, legal, and regulatory costs, including common language, common legal origin, and colonial ties. These variables are positively associated with the share of subsidiaries controlled by an intermediary in country k and, with the exception of common language, have similar effects in both delegation relationships and monitoring relationships. When the parent and middleman countries speak the same language, the share variable increases by 98\%, when it is the final country to speak the same language as the middleman, the increase is 55\%. Coefficients on shared predictors remain significant and quite stable between Columns (1) and (3). \\

Column (5) presents the full specification, which introduces the number of overlapping working hours between countries as a direct measure of frictions to real-time communication. This predictor has a significant impact on delegation and monitoring decisions. An increase of one overlapping working hour between the parent and intermediate affiliates increases the expected fraction of final subsidiaries controlled by the intermediate $k$ by 5.6\%. Similarly, an increase of one overlapping work hour between the intermediary and the location of the final subsidiary increases the expected share of subsidiaries controlled by the intermediate $k$ by 10.5\%. 

Importantly, the relative magnitudes of these effects match expectations. As anticipated, delegation costs are less binding than monitoring costs, meaning that a reduction in the second coefficient (monitoring) has about twice the effect of a reduction in the first (delegation). This is in line with our assumptions about Equation \ref{eq: composite cost}.

The inclusion of overlapping working hours also attenuates the estimated effects of geographical distance. Relative to Column (3), the distance coefficients decrease by nearly half, suggesting a positive correlation between distance and overlapping working hours. This pattern suggests that, in earlier specifications, geographical distance partially captures coordination frictions that are more directly measured by the overlapping-hours introduced in Column (5).

 \begin{table}[H]
\centering
\caption{Structural gravity results}
\label{tab:structural gravity}
\resizebox{.99\textwidth}{!}{%
\begin{tabular}{lcccccccc}
\hline
 & \\[-1.5ex]
 & (1) & (2) & & (3) & (4) & & (5) & (6) \\
  &  &  &  &  &  & & &\\
\textbf{Location of:} & \textbf{Middlemen} & \textbf{Final subsidiaries} & & \textbf{Middlemen} & \textbf{Final subsidiaries} & & \textbf{Middlemen} & \textbf{Final subsidiaries} \\ \cline{2-3} \cline{5-6}  \cline{8-9}\\
 & \\[-1.5ex]
Dep. var. &  $M_{ikj}/M_{ij}$ & $M_{ij}$ & & $M_{ikj}/M_{ij}$ & $M_{ij}$ & & $M_{ikj}/M_{ij}$ & $M_{ij}$\\[1ex] 
\cline{1-3}  \cline{5-6}  \cline{8-9}\\
 &  &  &  &  &  & & &\\

\textbf{N. of overlapping} & &  & &   & & &   \textbf{0.055***} & \\
\textbf{working hours$_{ik}$}& &  & & & & &  (0.006) & \\
 &  &  & & & & &  &\multicolumn{1}{l}{}   \\
\textbf{N. of overlapping} & &  & & & & & \textbf{0.100***} & \\
\textbf{working hours$_{kj}$} & &  & & & & &  (0.006) & \\
 &  &  & & &  & & &  \multicolumn{1}{l}{}  \\

$\text{Language}_{ik}$ & & & & 0.684*** & & & 0.685*** &\\
 & & & & (0.0318) & & & (0.0315) & \\
$\text{Language}_{kj}$ & & & & 0.441*** & & & 0.419*** & \\
 &  & & & (0.0326) & & & (0.0327) & \\
$\text{Colony dependence}_{ik}$ & & & & 1.513*** & & & 1.534*** & \\
 &  & & & (0.0483) & & & (0.0479) &\\
$\text{Colony dependence}_{kj}$ & & & & 1.503*** & & & 1.520*** & \\
 & & & & (0.0481) & & & (0.0479) & \\
$\text{Legal origins}_{ik}$ & & & & 0.231*** & & & 0.236*** & \\
 & & & & (0.0280) & & & (0.0280) & \\
$\text{Legal origins}_{kj}$ & & & & 0.221*** & & & 0.225*** &\\
 & & & & (0.0261) & & & (0.0262) & \\

 $\text{Distance (arcsin)}_{ik}$ (km) & --0.0575*** & & & --0.0615*** & & & --0.0302*** & \\
 & (0.00908) & & & (0.00922) & & & (0.00988) & \\
 $\text{Distance (arcsin)}_{kj}$ (km) & --0.306*** & & & --0.299*** & & & --0.172*** & \\
 & (0.0101) & & & (0.0101) & & & (0.0125) & \\
$\text{RTA}_{ik}$ & 1.408*** & & & 1.289*** & & & 1.261*** & \\
 & (0.0271) & & & (0.0275) & & & (0.0276) & \\
$\text{RTA}_{kj}$ & 0.930*** & & & 0.860*** & & & 0.807*** & \\
 & (0.0251) & & & (0.0257) & & & (0.0258) & \\
  $\text{Home}_{ik}$ & 5.319*** & & & 5.432*** & & & 5.392*** & \\
 & (0.0471) & & & (0.0494) & & & (0.0484) & \\
$\text{Home}_{kj}$ & 3.303*** & & & 3.438*** & & & 3.597*** & \\
 & (0.0596) & & & (0.0592) & & & (0.0571) & \\

 $\text{Direct control}_{ikj}$ & 0.696*** & & & 0.665*** & & & 0.836*** & \\
 & (0.0518) & & & (0.0528) & & & (0.0565) & \\

 $\sqrt{ \frac{\hat{\mathcal{C}}_{ij}}{K}}$  &  & --156.576*** & & & --154.541***& & & --140.892*** \\
 &  & (16.517) & & & (12.986) & & & (14.004) \\
 &  &  &  & & & & & \\ 
\cline{1-3}  \cline{5-6}  \cline{8-9}\\
  & \\[-1.5ex]
  Observations & 1,724,334 & 9,736 & & 1,718,840 & 9,712 & & 1,718,840 & 9,712\\
Fixed effects & $i\times j$ & i,j & & $i\times j$ & i,j & & $i\times j$ & i,j  \\[1ex] 
 &  &  & &  &  & & & \\
 \hline

\end{tabular}%
}
\begin{tablenotes}
 \footnotesize
     \item Note: Given the presence of zeros in the case of direct control, we scale the distances in km by using the hyperbolic sine transformation ($ln(x+\sqrt{x^2+1})$), which allows approximating to the natural logarithm while retaining the zeros \citep{bellemare2020elasticities}. Standard errors are double-clustered by origin and destination in parentheses (*** p$<$0.01, ** p$<$0.05, * p$<$0.1). 

\end{tablenotes}

\end{table}

\begin{figure}[H]
\centering
\caption{Distribution of the inspection cost} 
\label{fig: hat_c_hist}
\includegraphics[width=.6\linewidth]{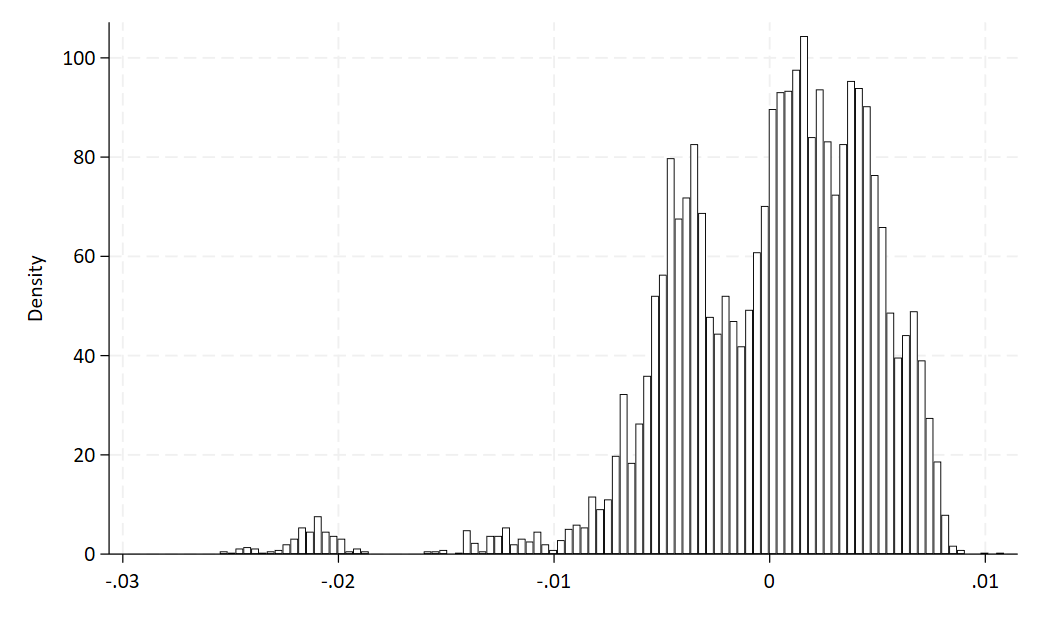}
\begin{tablenotes}
      \footnotesize 
      \item Note: The figure shows the distribution of the expected cost of corporate control estimated after Eq. \ref{eq:triangular structural}. The value can be negative or positive depending on the prevalence of gravity components (overlapping working hours, geographical distance, regional trade agreements, common language, colony dependence, legal origins) that proxy frictions with a negative or positive coefficient.
\end{tablenotes}
\end{figure}

Columns (2), (4) and (6) of Table \ref{tab:structural gravity} present the coefficients of the vector of the expected cost of corporate control ($E[C_{ikj}]$), obtained from previous estimates. The coefficients obtained correspond to $\theta$ in equation \ref{eq:bilateral structural}, and their sign and magnitude are as expected: an increase in the cost index reduces the number of subsidiaries in country $j$ controlled by parents in country $i$. Finally, Figure \ref{fig: hat_c_hist} shows the distribution of the expected inspection cost for each origin $i$ and destination $j$ obtained from the full model in column (5).

\section{Robustness and sensitivity checks}
\label{sec: RC}

In this section, we conduct several robustness checks to support our main conclusions. First, we examine whether tax optimization, rather than organizational costs, drives the location choice of intermediate jurisdictions. In the first column of Table \ref{tab:RC2}, we introduce controls for corporate tax differentials, calculated as the ratio of the average profit tax in the destination to the origin, using data from \cite{worldbank2020doingbusiness}. The profit tax measures the total burden on a typical firm as a percentage of commercial profits, and this is the same variable that has been used for preliminary evidence in Section \ref{sec: data}. These additional results confirm that investment shares in the $ikj$ triplet increase when country $k$ offers a relatively more favorable tax environment. However, while corporate tax differentials play a role, they do not diminish the importance of organizational costs in shaping ownership chains.

A second concern is whether labor costs, which are particularly important for multinational firms that coordinate global supply chains, are the most important determinant of the choice of the intermediate jurisdiction. In column 2, we extend our baseline model by incorporating labor cost differentials, measured as the ratio of employee compensation in the destination to that in the origin, using data from \cite{worldbank2020doingbusiness}. Not surprisingly, we find a negative and significant relationship between labor costs and investment decisions, but only in the bilateral relationship between countries $k$ and $i$. Higher labor costs in country $k$ discourage investment through that country, but the main findings on overlapping working hours remain significant, with estimated coefficients consistent in magnitude with the baseline results.

A third concern is that overlap in working hours may not be the only factor that facilitates communication. In column 3, we include an alternative proxy for ease of communication, the Common Language Index (CLI) proposed by \cite{gurevich2021one}. This index captures the linguistic similarity between countries, taking into account several dimensions of language, including translation ease and interpretability. While CLI has a strong positive impact on investment decisions, it does not eliminate the importance of overlapping working hours, confirming that real-time communication remains a distinct channel influencing delegation decisions.

In an additional set of robustness checks, we examine possible sample composition effects. Table \ref{tab: RC1} reports new estimates of the triangular equation across different industry groups. We split the results into three sub-samples: (i) parent companies in manufacturing (column 1), (ii) parent companies in services (excluding financial) (column 2), and (iii) cases where the parent, intermediary, and ultimate subsidiary are all in the financial sector (column 3).\footnote{A company's industry is identified using the 2017 NAICS classification at the 2-digit level. A firm is classified as financial if it is in "\textit{Finance and Insurance}" (code 52) or "\textit{Real Estate, Rental, and Leasing}" (code 53).}

\begin{table}[H]
\centering
\caption{Robustness: alternative stories}
\label{tab:RC2}
\resizebox{.7\textwidth}{!}{%
\begin{tabular}{lcccccc} \hline
& & & & & & \\[-2ex]

\textbf{Control for} & \multicolumn{2}{c}{Tax differentials} & \multicolumn{2}{c}{Wage differentials} & \multicolumn{2}{c}{Culture and language} \\

& & & & & & \\[-2ex]

\textbf{Dep Var.} & $M_{ikj}/M_{ij}$ & $M_{ij}$ & $M_{ikj}/M_{ij}$ & $M_{ij}$ & $M_{ikj}/M_{ij}$ & $M_{ij}$
\\[.5ex] \hline
& & & & & & \\
\textbf{N. of overlapping} & 0.087*** & & 0.088*** & & 0.052***  &\\
\textbf{working hours$_{ik}$} & (0.007) & & (0.008) & & (0.006) & \\
& & & & & & \\
\textbf{N. of overlapping} & 0.134*** & & 0.125*** & & 0.101*** & \\
\textbf{working hours$_{kj}$} & (0.008) & & (0.009) & & (0.006) & \\
& & & & & & \\
$\frac{CT_{k}}{CT_{i}}$ & 0.001 & &  &  & & \\
 & (0.001) &  & & & &  \\
$\frac{CT_{j}}{CT_{k}}$ & 0.008*** &  & & & &  \\
 & (0.000) &  &  & & & \\
$\frac{LC_{k}}{LC_{i}}$ &  & & -0.638*** & & &  \\
 &  & & (0.034) &  & &  \\
$\frac{LC_{j}}{LC_{k}}$ & & & -0.049* &  & & \\
 & & & (0.029) &  & &  \\
CLI$_{ik}$ &  &  & & & 1.519*** & \\
 &  &  & & & (0.057) & \\
CLI$_{kj}$ &  &  & & & 1.187*** & \\
 &  &  & & & (0.056) & \\
Distance (arcsin)$_{ik}$ (km) & -0.125*** & & -0.124*** & & -0.00633 & \\
 & (0.012) & & (0.015) & & (0.010) & \\
Distance (arcsin)$_{kj}$ (km) & -0.201*** & & -0.207*** & & -0.137*** & \\
 & (0.015) & & (0.018) & & (0.013)& \\
$\text{RTA}_{ik}$ & 1.204*** & & 1.080*** & & 1.217*** & \\
 & (0.033) & & (0.041) & & (0.028) & \\
$\text{RTA}_{kj}$ & 0.797*** & & 0.681*** & & 0.731*** & \\
 & (0.030) & & (0.037) & & (0.027) & \\
$Home_{ik}$ & 5.168*** & & 4.693*** & & 4.453*** & \\
 & (0.055) & & (0.064) & & (0.054) &\\
$Home_{kj}$ & 3.587*** & & 3.354*** & & 2.906*** & \\
 & (0.065) & & (0.070) & & (0.060) & \\
Language$_{ik}$ & 0.659*** & & 0.809*** & & & \\
 & (0.0359) & & (0.0412) & & &  \\
Language$_{kj}$ & 0.380*** & & 0.550*** &  && \\
 & (0.038) & & (0.044) & & &  \\
Colony dependence$_{ik}$ & 1.674*** & & 1.590*** & & 1.526*** & \\
 & (0.051) & & (0.055) & & (0.048) & \\
Colony dependence$_{kj}$ & 1.631*** & & 1.386*** & & 1.431*** & \\
 & (0.052) & & (0.059) & & (0.048) & \\
Legal origins$_{ik}$ & 0.166*** & & 0.222*** & & 0.124*** & \\
 & (0.032) & & (0.035) & & (0.029) & \\
Legal origins$_{kj}$ & 0.190*** & & 0.308*** & & 0.145*** & \\
 & (0.030) & & (0.033) & & (0.027) & \\
Direct$_{ikj}$ & 0.220*** & & 0.156* & & 0.966*** & \\
 & (0.073) & & (0.088) & & (0.056) & \\
$\sqrt{ \frac{\hat{\mathcal{C}}_{ij}}{K}}$  & & -135.109*** &  & -135.218*** &  & -141.157***\\
 &  & (16.139) &  & (17.558) & & (13.547)\\ 
 &  &  & & & &  \\ \hline
Observations & 1,224,038 & 8,000 & 641,648 & 5,386 & 1,718,840 & 9,712 \\
Fixed effects & $i\times j$ & i,j & $i\times j$ & i,j & $i\times j$ & i,j \\ \hline 
\end{tabular}}
\begin{tablenotes}
    \footnotesize
    \item Note: Ratios $CT_{k} / CT_{i}$ and $CT_{j} / CT_{k}$ control for relative advantage in corporate taxation. Ratios $LC_{k} / LC_{i}$ and $LC_{j} / LC_{k}$ control for the relative advantage in labor costs. Given the presence of zeros in the case of direct control, we scale the distances in km by using the hyperbolic sine transformation ($ln(x+\sqrt{x^2+1})$), which allows approximating to the natural logarithm while retaining the zeros \citep{bellemare2020elasticities}. Standard errors are double-clustered by origin and destination in parentheses (*** p$<$0.01, ** p$<$0.05, * p$<$0.1). 

\end{tablenotes}
\end{table}

\begin{table}[H]
\centering
\caption{Sensitivity to sample composition}
\label{tab: RC1}
\resizebox{.8\textwidth}{!}{%
\begin{tabular}{lcccccc} \hline
& & & & & & \\[-2ex]
\textbf{Sample} & \multicolumn{2}{c}{Manufacturing} & \multicolumn{2}{c}{Service} & \multicolumn{2}{c}{Finance} \\
& & & & & & \\[-2ex]
\textbf{Dep Var.} & $M_{ikj}/M_{ij}$ & $M_{ij}$ & $M_{ikj}/M_{ij}$ & $M_{ij}$ & $M_{ikj}/M_{ij}$ & $M_{ij}$
\\[.5ex] \hline
& & & & & & \\
\textbf{N. of overlapping} & 0.023*** & & 0.059*** & & 0.141*** & \\
\textbf{working hours$_{ik}$} & (0.008) & & (0.008) & & (0.024) & \\
& & & & & & \\[-0.5ex]
\textbf{N. of overlapping} & 0.108*** & & 0.102*** & & 0.176*** & \\
\textbf{working hours$_{kj}$} & (0.009) & & (0.008) & & (0.024) & \\
& & & & & & \\
Distance (arcsin)$_{ik}$ & -0.172*** & & -0.085*** & & 0.024 & \\
 & (0.017) & & (0.012) & & (0.029) & \\
Distance (arcsin)$_{kj}$  & -0.200*** & & -0.200*** & & -0.010 & \\
 & (0.020) & & (0.016) & & (0.032) & \\
RTA$_{ik}$ & 1.176*** & & 1.279*** & & 1.212*** & \\
 & (0.042) & & (0.037) & & (0.086) & \\
RTA$_{kj}$ & 0.874*** & & 0.840*** & & 0.954*** &\\
 & (0.038) & & (0.034) & & (0.084) & \\
Home$_{ik}$ & 4.702*** & & 5.586*** & & 5.788*** &\\
 & (0.078) & & (0.062) & & (0.142) & \\
Home$_{kj}$ & 3.408*** & & 3.567*** & & 5.380*** & \\
 & (0.089) & & (0.074) & & (0.157) & \\
Language$_{ik}$ & 0.594*** & & 0.855*** & & 0.546*** & \\
 & (0.0450) & & (0.0414) & & (0.102) & \\
Language$_{kj}$ & 0.274*** & & 0.497*** & & 0.596***  &\\
 & (0.046) & & (0.044) & & (0.107) & \\
Colony dependence$_{ik}$ & 1.087*** & & 1.483*** & & 1.008*** & \\
 & (0.080) & & (0.064) & & (0.157) & \\
Colony dependence$_{ik}$ & 1.342*** & & 1.614*** & & 1.222*** & \\
 & (0.081) & & (0.063) & & (0.157) &\\
Legal origins$_{ik}$ & 0.196*** & & 0.150*** & & 0.371*** &\\
 & (0.040) & & (0.037) & & (0.100) & \\
Legal origins$_{kj}$ & 0.268*** & & 0.129*** & & 0.220** & \\
 & (0.039) & & (0.034) & & (0.098) & \\
Direct$_{ikj}$ & 0.434*** & & 0.158** & & 0.860*** & \\
 & (0.095) & & (0.072) & & (0.169) & \\
$\sqrt{ \frac{\hat{\mathcal{C}}_{ij}}{N_K}}$  & & -89.573*** &  & -158.517*** &  & -118.742*** \\
 &  & (21.185) &  & (12.226) & & (7.564)\\ 
 &  &  &  &  &  & \\ \hline
 &  &  &  &  &  &  \\[-2ex]
Observations & 836,425 & 4,715 & 1,083,901 & 6,110 & 301,731 & 1,671 \\
Fixed effects & $i\times j$ & i,j & $i\times j$ & i,j & $i\times j$ & i,j \\ \hline
\end{tabular}}
\begin{tablenotes}
    \footnotesize
    \item Note: Sectors are defined according to the NAICS 2017 classification at the 2-digit level. Given the presence of zeros in the case of direct control, we scale the distances in km by using the hyperbolic sine transformation ($ln(x+\sqrt{x^2+1})$), which allows approximating to the natural logarithm while retaining the zeros \citep{bellemare2020elasticities}. Standard errors are double-clustered by origin and destination in parentheses (*** p$<$0.01, ** p$<$0.05, * p$<$0.1). 

\end{tablenotes}
\end{table}

In particular, there is an increase in the gap between the first and second coefficients in the manufacturing sector with respect to the baseline results. In the services sector, this gap remains similar to the baseline, while it disappears completely in the financial groups. This pattern is in line with our theoretical framework. The difference between the costs of delegating and monitoring is expected to be most pronounced in manufacturing, where coordination often involves complex but fragmented global supply chains. In contrast, the contractual environment of financial firms differs from that of manufacturing and service firms, even though they also require cross-border communication and delegation. This difference is particularly relevant for firms operating in publicly traded markets, where financial transactions must be coordinated with the opening and closing hours of markets in different time zones.

A final concern is whether ownership chains are systematically structured around tax havens. The global coverage of our database allows us to quantify the prevalence of intermediate subsidiaries in recognized tax havens, based on the classifications of \cite{hines1994fiscal, hines2010treasure}. Across all parent and subsidiary locations, we find that 7.4\% of intermediate affiliates are located in tax havens. While this share is not negligible, it does not imply that tax havens are fully determinants of the ownership chain structure.

\section{Conclusions}
\label{sec: concl}

This study examines how multinational enterprises (MNEs) structure complex hierarchical ownership networks, where control is often exercised through multi-tiered chains of subsidiaries spanning multiple national borders. Although these structures play a crucial role in the global organization of MNEs, they have been largely overlooked in the literature, likely due to data limitations. Using a unique global database, we first document stylized facts on the geography of ownership chains and then develop a theoretical framework based on the idea that communication frictions among managers play a crucial role.


When we examine the geographic organization of complex ownership chains, we find they can be long and strech accross multiple country borders. Approximately 61\% of subsidiaries are positioned at least two ownership links’ distance from the parent company, 47\% cross at least two country borders, and 53\% extend approximately more than 10,000 kilometers.

In particular, we find that there are two relevant correlations: i) between the hierarchical position of subsidiaries and their geographic distance from headquarters; ii) between the hierarchical position of subsidiaries and the difference in time zones, i.e, the overlapping in working hours. The latter is for us a measure of the ease of communication among managers across countries. Briefly, subsidiaries at the bottom of the hierarchy share fewer working hours with managers at the top.

Building on these empirical patterns, we propose a location choice model in which parent firms compete in a global auction to acquire new subsidiaries. In formulating their bids, parents weigh the option of delegating monitoring responsibilities to an intermediate affiliate in a third country. The model yields two testable equations: the first is a trilateral equation that determines whether monitoring is delegated to an intermediate affiliate and where it is located, while the second is a bilateral equation that predicts the location of newly acquired final subsidiaries.

Crucially, we find that communication frictions are a key determinant of delegation and monitoring decisions. Even after subjecting our results to extensive robustness checks, our main findings remain intact. The impact of communication costs persists even after controlling for alternative explanations such as tax differentials, tax haven status, wage differentials, and cultural/linguistic proximity.

Our results highlight the limited understanding of the ownership structures of multinational firms, despite their significant weight in the global economy. While official statistics often fail to fully capture global ownership networks, the increasing standardization of national business registries is improving data availability and accuracy. Looking ahead, we see promising opportunities for future research on the emergence and dynamics of multinational ownership chains.

\setlength\bibsep{0.5pt}
\footnotesize
\bibliographystyle{elsarticle-harv}
\bibliography{bibliography.bib}

\newpage

\appendix
\section*{Appendix A: Ownership chains and corporate control networks}
\label{sec: appendix a}

\setcounter{table}{0}
\renewcommand{\thetable}{A\arabic{table}}
\setcounter{figure}{0}
\renewcommand{\thefigure}{A\arabic{figure}}

\begin{figure}[H]
    \centering
    \caption{A fictional corporate tree of an MNE}
     \resizebox{0.35\textwidth}{!}{%
\includegraphics[width=\textwidth]{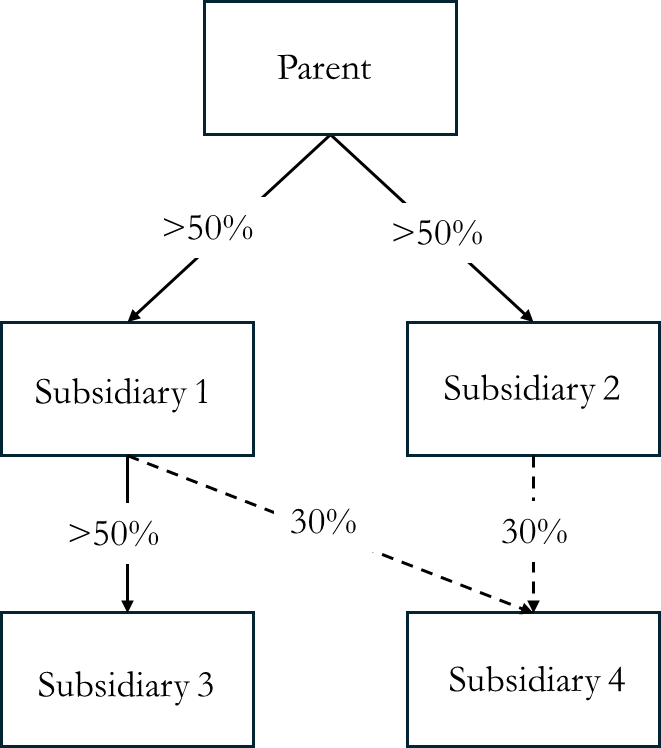}}
   \label{fig: fictional tree}
      \begin{tablenotes}
      \footnotesize
\item Note: The figure represents a fictional corporate tree where Subsidiaries 1 and 2 are directly controlled, Subsidiary 3 is indirectly controlled by transitivity, and Subsidiary 4 is controlled by consolidation of fragmented equity stakes.
\end{tablenotes}
\end{figure}

In this Appendix, we provide an understanding of how raw shareholding data are used to extract corporate trees developed by MNEs. For further details, we refer to \cite{rungi2017global}. An MNE is made of a unique parent company and its subsidiaries. If we look at their ownership data, they represent a corporate network in the form of a hierarchical tree, where nodes are legally autonomous companies. Please consider the simplified fictional corporate tree in Figure \ref{fig: fictional tree}. On top of the hierarchy, we find the parent company, which exerts control over subsidiaries by means of direct and indirect shareholding links. In principle, an MNE can develop much more complex patterns than the ones depicted in Figure \ref{fig: fictional tree}, as it can include cross-holdings when shareholding relationships run two-way between companies. They can also entail ownership cycles, i.e. when the firm holds a share of its equity stakes. Given a fragmentation of shareholding, companies can indirectly connect through multiple sequences of ownership links.  The solution is to start with a full ownership matrix of all firms and their shareholders in the data. Thus, a control relationship is detected iteratively on a transformation of the ownership matrix. In line with international accounting standards \citep{oecd2005mne, unctad2009fdi, eurostat2007for}, one can assume that corporate control is present when there is an absolute majority of equity stakes. Yet, one cannot exclude that lower thresholds still allow the parent company the ability to influence and contribute to the decision-making of another company.

In Figure \ref{fig: fictional tree}, we easily attribute direct control of Subsidiary 1 to the Parent because the latter holds an absolute majority of voting rights in the first. Besides this most elementary case of direct control, there are two other ways for the Parent to exert control: (i) \textit{by transitivity}, when successive majority links form an ownership chain, and subsidiaries align on it, as in the case of Subsidiary 3; (ii) \textit{by consolidation}, when equity is fragmented, no direct majority path can be detected from the Parent, and yet other subsidiaries hold minority links in a firm, like Subsidiary 4, which can be consolidated and allow the unique Parent to finally reach out to the management of Subsidiary 4.\\ 

\textbf{Ownership chains.} Figure \ref{fig: fictional tree} shows an MNE that has two ownership chains. One starts from the Parent and reaches out to Subsidiary 3 through Subsidiary 1. A second one starts from the Parent and reaches out to Subsidiary 4 thanks to either Subsidiary 1 or Subsidiary 3. To solve uncertainty, we will assume that the final subsidiary is on the shortest ownership chain. If ownership chains are of equal length, then we discriminate on shares. When shares are equal, we discriminate by firm size, giving priority to the middleman that is bigger.\\

\textbf{Middlemen subsidiaries.} As evident from Figure \ref{fig: fictional tree}, we have a category of subsidiaries that is peculiar because they are controlled by the Parent, but they also hold controlling stakes in other subsidiaries. Economic literature has neglected this category of subsidiaries. We call them middlemen, with reference to their intermediary role between the management of the Parent and the management of the downstream subsidiary. Middlemen subsidiaries can present themselves in a sequence over longer ownership chains.

\newpage

\appendix

\newpage
\appendix
\section*{Appendix B: Tables and graphs }
\label{sec: appendix b}

\setcounter{table}{0}
\renewcommand{\thetable}{B\arabic{table}}
\setcounter{figure}{0}
\renewcommand{\thefigure}{B\arabic{figure}}

\begin{table}[H]
\centering
\caption{Country-level covariates and sources}
\label{tab: country covariates}
\resizebox{\textwidth}{!}{%
\begin{tabular}{lll}
\hline \hline 
& & \\[-2ex]
\textbf{Variable} & \textbf{Description} & \textbf{Source} \\[1ex] \hline
& & \\[-2ex]
\begin{tabular}[c]{@{}l@{}}Number of overlapping \\ working hours\end{tabular} & \begin{tabular}[c]{@{}l@{}} An integer variable counting the number of working hours during which offices in two countries \\ are simultaneously open. Assuming a standard 10-hour workday, this is calculated by subtracting \\ the time zone difference (in hours) between the origin and destination countries from the total of \\ ten hours.\end{tabular} & \cite{bahar2020hardships} \\
& & \\
Total Tax Contribution Rate (TTCR) &  \begin{tabular}[c]{@{}l@{}} Total amount of taxes and mandatory contributions payable by businesses, as a percentage \\of commercial profit. \end{tabular}  & \cite{worldbank2020doingbusiness} \\
& & \\
Corporate Taxation (CT) &  \begin{tabular}[c]{@{}l@{}}	Profit tax calculated by the World Bank as the total amount of taxes paid by the business as a \\ percentage of commercial profits. \end{tabular}  & \cite{worldbank2020doingbusiness} \\
 &  & \\
Cost of Labour (CL) & \begin{tabular}[c]{@{}l@{}} Compensation of employees calculated by the World Bank as the total of cash payments, as well \\ 
as in kind (such as food and housing), to employees in return for services rendered, and government \\ contributions to social insurance schemes such as social security and pensions that provide benefits \\ to employees. \end{tabular} & \cite{worldbank2020doingbusiness} \\
& & \\
Common Language Index (CLI) & \begin{tabular}[c]{@{}l@{}} A continuous measure of linguistic similarity both within and between countries, that captures \\ three aspects: shared official languages, common native languages, and linguistic proximity. \end{tabular} & \cite{gurevich2021one}\\
 & & \\[1ex] \hline
\end{tabular}%
}
\end{table}

\begin{table}[H]
    \centering
    \caption{Sample coverage by industry}
         \resizebox{0.99\textwidth}{!}{%
    \begin{tabular}{lclcccc}
    \hline
        ISIC & NACE & Industry description & N. subsidiaries & \% & N. middlemen & \% \\ \hline
        A & 01 -03 & Agriculture, forestry and fishing & 11,967 & 0.94\% & 1,846 & 0.69\% \\ 
        B & 05 - 09 & Mining and quarrying & 17,969 & 1.41\% & 4556 & 1.71\% \\ 
        C & 10 - 33 & Manufacturing & 239,091 & 18.74\% & 53,809 & 20.19\% \\ 
        D & 35 &  Electricity, gas, steam and air conditioning supply & 33,982 & 2.66\% & 4,235 & 1.59\% \\ 
        E & 36-39 &  Water supply; sewerage, waste management & 8,728 & 0.68\% & 1,477 & 0.55\% \\ 
        F & 41 - 43 &  Construction & 52,971 & 4.15\% & 8,750 & 3.28\% \\ 
        G & 45 - 47 & Wholesale and retail trade  & 207,516 & 16.27\% & 30,182 & 11.33\% \\ 
        H & 49 - 53 & Transportation and storage & 48,223 & 3.78\% & 7,698 & 2.89\% \\ 
        I & 55 - 56 & Accommodation and food service activities & 23,366 & 1.83\% & 2,982 & 1.12\% \\ 
        J & 58 - 63 &  Information and communication & 80,985 & 6.35\% & 16,618 & 6.24\% \\ 
        K & 64 - 66 & Financial and insurance activities & 187,421 & 14.69\% & 63,538 & 23.84\% \\ 
        L & 68 & Real estate activities & 87,961 & 6.89\% & 12,205 & 4.58\% \\ 
        M & 69 - 75 & Professional, scientific and technical activities & 130,179 & 10.20\% & 33,222 & 12.47\% \\ 
        N & 77 - 82 & Administrative and support service activities & 87,332 & 6.85\% & 16,803 & 6.31\% \\ 
        O & 84 & Public administration and defense  & 835 & 0.07\% & 212 & 0.08\% \\ 
        P & 85 &  Education  & 6,542 & 0.51\% & 953 & 0.36\% \\ 
        Q & 86 - 88 & Human health and social work activities & 22,795 & 1.79\% & 3,289 & 1.23\% \\ 
        R & 90 - 93 &  Arts, entertainment and recreation & 9,292 & 0.73\% & 1551 & 0.58\% \\ 
        S & 94 - 96 & Other service activities & 16,418 & 1.29\% & 2,236 & 0.84\% \\ 
        T & 97 - 98 & Activities of households as employers; etc.  & 2,044 & 0.16\% & 278 & 0.10\% \\ 
        U & 99 & Activities of extraterritorial organizations & 149 & 0.01\% & 25 & 0.01\% \\ \hline
        ~ & ~ & Total & 1,275,766 & 100.00\% & 266,465 & 100.00\% \\ \hline
    \end{tabular}}
    \label{mne industry}
    \begin{tablenotes}
    \footnotesize
        \item Note: We report the industry coverage of our sample by adopting the main aggregates of the NACE rev. 2 and ISIC rev. 4 classifications. First, we report the salience of subsidiaries (columns 4 and 5) and then we separate among them the number of middlemen subsidiaries (columns 6 and 7).
    \end{tablenotes}
\end{table}

\begin{table}[!ht]
    \centering
    \caption{Sample coverage by size}
    \resizebox{0.6\textwidth}{!}{%
    \begin{tabular}{lcccc}
    \hline
        Size category & N. subsidiaries & \% & N. middlemen & \% \\ \hline
        Small & 1,205,973 & 67.33\% & 127,535 & 40.85\% \\ 
        Medium & 283,541 & 15.83\% & 50,063 & 16.04\% \\ 
        Large & 204,956 & 11.44\% & 63,563 & 20.36\% \\ 
        Very large & 96,543 & 5.39\% & 71,048 & 22.76\% \\ \hline 
        Total & 1,791,013 & 100.00\% & 312,209 & 100.00\% \\ \hline
    \end{tabular}}
    \label{mne size}
    \begin{tablenotes}
    \footnotesize
       \item Note: The table presents sample coverage of all subsidiaries and, among them, middlemen subsidiaries based on a combination of criteria: A) Large or very large companies report more than 10 million EUR revenues, or more than 20 million EUR total assets, or more than 150 employees, or over 0.5 million EUR capitalization, or they are listed at a stock exchange; B) Medium-sized companies register more than 1 million EUR revenues, or more than 2 million EUR total assets, or more than 15 employees, or over 50 thousand EUR capitalization; C) Small companies includes the residual, i.e. they are not in the medium or the large and very large categories. 
    \end{tablenotes}
\end{table}

\begin{figure}[H]
    \centering
    \caption{Financial accounts - geographic coverage in 2019}
     \resizebox{\textwidth}{!}{%
\includegraphics[width=\textwidth]{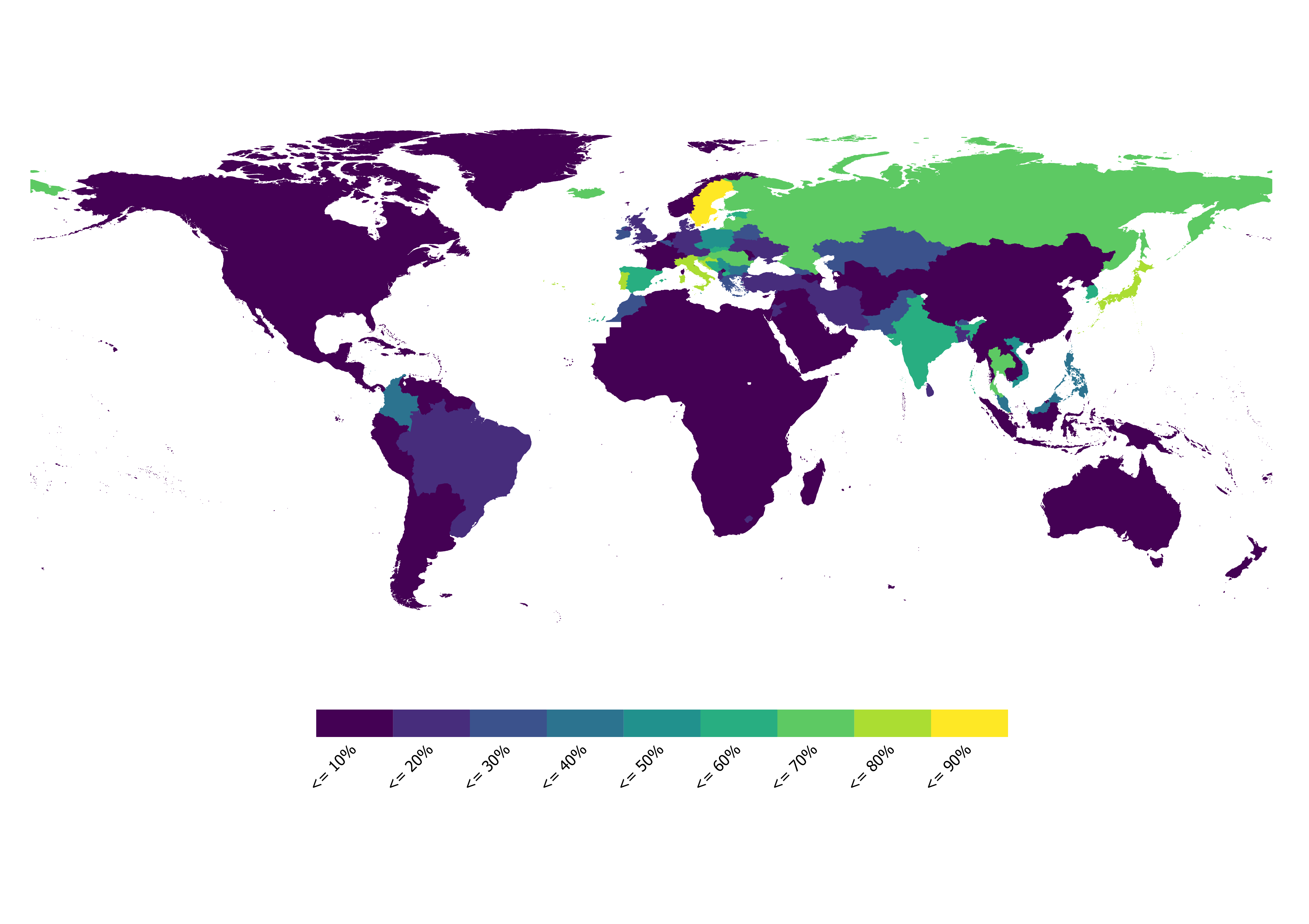}}
   \label{fig: nonmiss_sh}
      \begin{tablenotes}
      \footnotesize
\item Note: The map displays the sample coverage of financial accounts in different countries. Percentages indicate the availability of basic information on operating revenues/turnover for the total number of firms for which ownership data are available.
\end{tablenotes}
\end{figure}

\begin{figure}[H]
\centering
\caption{Size distribution - (log of) number of controlled subsidiaries}
\label{fig: size density}
\includegraphics[width=0.7\linewidth]{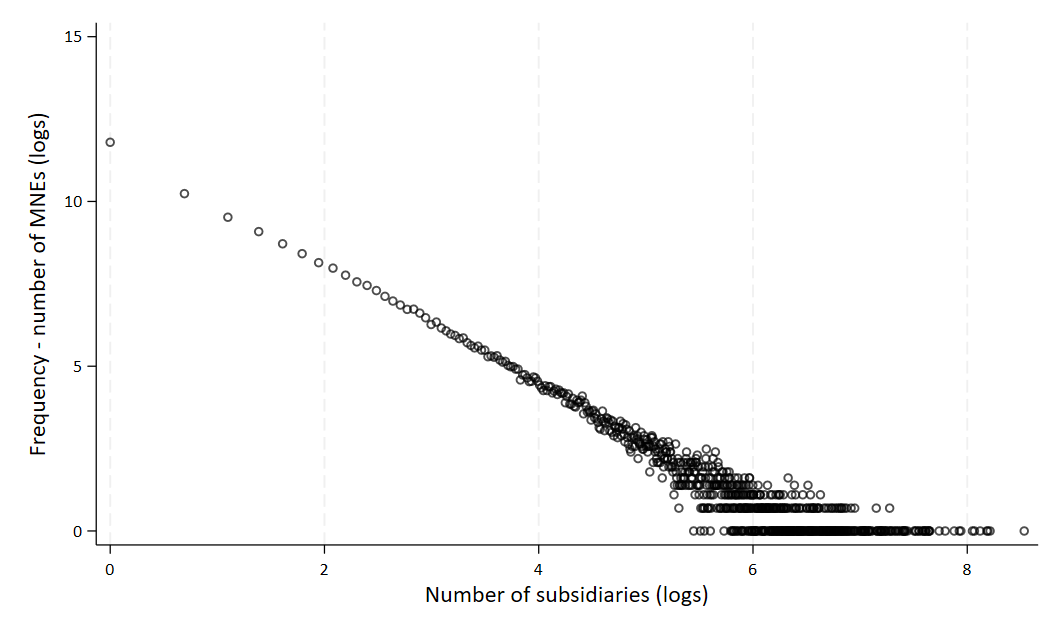}
\begin{tablenotes}
      \scriptsize 
      \item Note: The size of MNEs is measured as the logarithm of the number of controlled subsidiaries. By assuming that a negative binomial distribution can represent the data-generating process, we obtain an overdispersion parameter of $0.58$ and a mean of $0.63$.
\end{tablenotes}
\end{figure}

\begin{table}[H]
\centering
\caption{Lengths of ownership chains}
\label{tab: complex control chains}
\resizebox{\textwidth}{!}{%
\begin{tabular}{lc|cccccccc|l}
\cline{3-10}
\multicolumn{1}{c}{} & \multicolumn{1}{l|}{} & \multicolumn{8}{c|}{\textbf{N. of country borders crossed}} &  \\ \hline
\multicolumn{2}{l|}{\textbf{N. of subsidiaries}} & \multicolumn{1}{c|}{Domestic} & 1 & 2 & 3 & 4 & 5 & 6 & 7 & Total \\ \hline
 & \\[-1.5ex]
\begin{tabular}[c]{@{}l@{}}Simple \\ MENs\end{tabular} & 1 & \multicolumn{1}{c|}{51,680} & 222,186 &  &  &  &  &  &  & 273,866 \\[-1.8ex]
 &  & \multicolumn{1}{c|}{{\footnotesize (3.406\%)}} & {\footnotesize (14.645\%)} &  &  &  &  &  &  & {\footnotesize (18.051\%)} \\ \hline
 & 1 & \multicolumn{1}{c|}{223,995} & 191,138 &  &  &  &  &  &  & 415,133 \\[-.5ex]
 &  & \multicolumn{1}{c|}{{\footnotesize (14.764\%)}} & {\footnotesize (12.599\%)} &  &  &  &  &  &  & {\footnotesize (27.363\%)} \\ 
 & 2 & \multicolumn{1}{c|}{181,428} & \cellcolor[HTML]{EFEFEF}174,195 & \cellcolor[HTML]{EFEFEF}41,266 &  &  &  &  &  & 396,889 \\[-.5ex]
 &  & \multicolumn{1}{c|}{{\footnotesize (11.959\%)}} & \cellcolor[HTML]{EFEFEF} {\footnotesize (11.482\%)} & \cellcolor[HTML]{EFEFEF} {\footnotesize (2.720\%)} &  &  &  &  &  & {\footnotesize (26.160\%)} \\
 & 3 & \multicolumn{1}{c|}{94,778} & \cellcolor[HTML]{EFEFEF}80,172 & \cellcolor[HTML]{EFEFEF}30,602 & \cellcolor[HTML]{EFEFEF}5,522 &  &  &  &  & 211,074 \\[-.5ex]
Complex &  & \multicolumn{1}{c|}{{\footnotesize (6.247\%)}} & \cellcolor[HTML]{EFEFEF} {\footnotesize (5.284\%)} & \cellcolor[HTML]{EFEFEF} {\footnotesize (2.017\%)} &  \cellcolor[HTML]{EFEFEF} {\footnotesize (0.364\%)} &  &  &  &  & {\footnotesize (13.913\%)} \\
MNEs & 4 & \multicolumn{1}{c|}{42,333} & \cellcolor[HTML]{EFEFEF}37,480 & \cellcolor[HTML]{EFEFEF}19,856 & \cellcolor[HTML]{EFEFEF}6,037 & \cellcolor[HTML]{EFEFEF}1,081 &  &  &  & 106,787 \\[-.5ex]
 &  & \multicolumn{1}{c|}{{\footnotesize (2.790\%)}} & \cellcolor[HTML]{EFEFEF} {\footnotesize (2.470\%)} & \cellcolor[HTML]{EFEFEF} {\footnotesize (1.309\%)} &  \cellcolor[HTML]{EFEFEF} {\footnotesize (0.398\%)} & \cellcolor[HTML]{EFEFEF} {\footnotesize (0.071\%)} &  &  &  & {\footnotesize (7.039\%)} \\ 
 & 5 & \multicolumn{1}{c|}{17,663} & \cellcolor[HTML]{EFEFEF}18,363 & \cellcolor[HTML]{EFEFEF}11,407 & \cellcolor[HTML]{EFEFEF}5,189 & \cellcolor[HTML]{EFEFEF}961 & \cellcolor[HTML]{EFEFEF}205 &  &  & 53,788 \\[-.5ex]
 &  & \multicolumn{1}{c|}{{\footnotesize (1.164\%)}} & \cellcolor[HTML]{EFEFEF} {\footnotesize (1.210\%)} & \cellcolor[HTML]{EFEFEF} {\footnotesize (0.752\%)} &  \cellcolor[HTML]{EFEFEF} {\footnotesize (0.342\%)} & \cellcolor[HTML]{EFEFEF} {\footnotesize (0.063\%)} & \cellcolor[HTML]{EFEFEF} {\footnotesize (0.014\%)}  &  &  & {\footnotesize (3.545\%)} \\
 & 6 & \multicolumn{1}{c|}{7,177} & \cellcolor[HTML]{EFEFEF}8,010 & \cellcolor[HTML]{EFEFEF}6,558 & \cellcolor[HTML]{EFEFEF}3,865 & \cellcolor[HTML]{EFEFEF}1,014 & \cellcolor[HTML]{EFEFEF}203 & \cellcolor[HTML]{EFEFEF}7 &  & 26,834 \\[-.5ex]
 &  & \multicolumn{1}{c|}{{\footnotesize (0.473\%)}} & \cellcolor[HTML]{EFEFEF} {\footnotesize (0.528\%)} & \cellcolor[HTML]{EFEFEF} {\footnotesize (0.432\%)} &  \cellcolor[HTML]{EFEFEF} {\footnotesize (0.255\%)} & \cellcolor[HTML]{EFEFEF} {\footnotesize (0.067\%)} & \cellcolor[HTML]{EFEFEF} {\footnotesize (0.013\%)}  & \cellcolor[HTML]{EFEFEF} {\footnotesize (0.000\%)} &  & {\footnotesize (1.769\%)} \\
 & \textgreater{}=7 & \multicolumn{1}{c|}{5,595} & \cellcolor[HTML]{EFEFEF}8,375 & \cellcolor[HTML]{EFEFEF}8,622 & \cellcolor[HTML]{EFEFEF}6,273 & \cellcolor[HTML]{EFEFEF}2,869 & \cellcolor[HTML]{EFEFEF}810 & \cellcolor[HTML]{EFEFEF}196 & \cellcolor[HTML]{EFEFEF}27 & 32,767 \\[-.5ex]
 &  & \multicolumn{1}{c|}{{\footnotesize (0.369\%)}} & \cellcolor[HTML]{EFEFEF} {\footnotesize (0.552\%)} & \cellcolor[HTML]{EFEFEF} {\footnotesize (0.568\%)} &  \cellcolor[HTML]{EFEFEF} {\footnotesize (0.413\%)} & \cellcolor[HTML]{EFEFEF} {\footnotesize (0.189\%)} & \cellcolor[HTML]{EFEFEF} {\footnotesize (0.053\%)}  & \cellcolor[HTML]{EFEFEF} {\footnotesize (0.013\%)} & \cellcolor[HTML]{EFEFEF} {\footnotesize (0.002\%)} & {\footnotesize (2.160\%)} \\ \hline 
\multicolumn{2}{l|}{Total} & \multicolumn{1}{c|}{624,649} & 739,919 & 118,311 & 26,886 & 5,925 & 1,218 & 203 & 27 & 1,517,138 \\[-.8ex]
 &  & \multicolumn{1}{c|}{{\footnotesize (41.173\%)}} & {\footnotesize (48.771\%)} & {\footnotesize (7.798\%)} & {\footnotesize (1.772\%)} & {\footnotesize (0.391\%)} & {\footnotesize (0.080\%)}  & {\footnotesize (0.013\%)} & {\footnotesize (0.002\%)} & {\footnotesize (100.000\%)} \\ 
\end{tabular}%
}
\begin{tablenotes}
 \scriptsize 
\item Note: The observation unit of this table is the unique ownership chain running from the parent to a subsidiary, therefore including the middlemen and the third jurisdictions crossed on the way. The y-axis indicates the length of the hierarchy, while the x-axis indicates the number of countries encountered along the chain. In row 1, we separate simpler MNEs whose subsidiaries are foreign or domestic but are all controlled by direct links, i.e., no middleman exists. The first column reports domestic subsidiaries by MNEs at varying lengths, where at least one middleman exists. In the grey area, we highlight ownership chains crossing more than one country where middlemen can be encountered.   
\end{tablenotes}
\end{table}

\section*{Appendix C: Additional stylized facts on multinational enterprises and their corporate structures}
\label{sec: appendix c}

\setcounter{table}{0}
\renewcommand{\thetable}{C\arabic{table}}
\setcounter{figure}{0}
\renewcommand{\thefigure}{C\arabic{figure}}

The role of indirect control relationships and ownership chains in MNEs has been largely overlooked in the economic literature, most likely due to data limitations. To fill this gap, we first present some stylized facts about complex ownership chains. Then, we turn to the theoretical and empirical framework that explains why such chains can arise.

\begin{itemize}
    \item \textsl{Most subsidiaries of multinational enterprises are indirectly controlled through ownership chains.} 
\end{itemize}

\noindent A key observation is that most subsidiaries of MNEs are controlled indirectly through ownership chains. Table \ref{tab: type of control} classifies all subsidiaries in our dataset based on the type of control exercised by the parent companies, regardless of geographic location or firm size. We find that 54\% of subsidiaries are indirectly controlled, underscoring the prevalence of multi-tiered corporate structures. Referring to the hypothetical ownership chain in Figure \ref{fig: fictional chain}, directly controlled subsidiaries maintain a direct link to the parent company, while indirectly controlled subsidiaries are linked through other subsidiaries, forming multilevel ownership chains.

\begin{table}[H]
    \centering
    \caption{Direct and indirect control of subsidiaries}
    \resizebox{.4\textwidth}{!}{%
    \begin{tabular}{lcc}
    \hline
        Type of control & N. subsidiaries & \% \\ \hline
        Direct control & 827,516 & 45.81 \\ 
        Indirect control & 978,952 & 54.19 \\ 
        Total & 1,806,468 & 100.00 \\ \hline
    \end{tabular}}
    \label{tab: type of control}
    \begin{tablenotes}
        \footnotesize
        \item Note: a parent company exerts direct control over a subsidiary when it holds the majority of voting rights. Indirect control occurs when the parent company controls a subsidiary through an intermediate affiliate within an ownership chain.
    \end{tablenotes}
\end{table}

\begin{itemize}
    \item \textsl{Complex multinational enterprises with ownership chains have a relevant economic weight, as they represent the lion's share (95\%) of global sales by multinational enterprises.} 
\end{itemize}

\begin{figure}[H]
\centering
\caption{MNEs, ownership chains, and their economic weight}
\label{fig: economic weight}
\includegraphics[width=.7\linewidth]{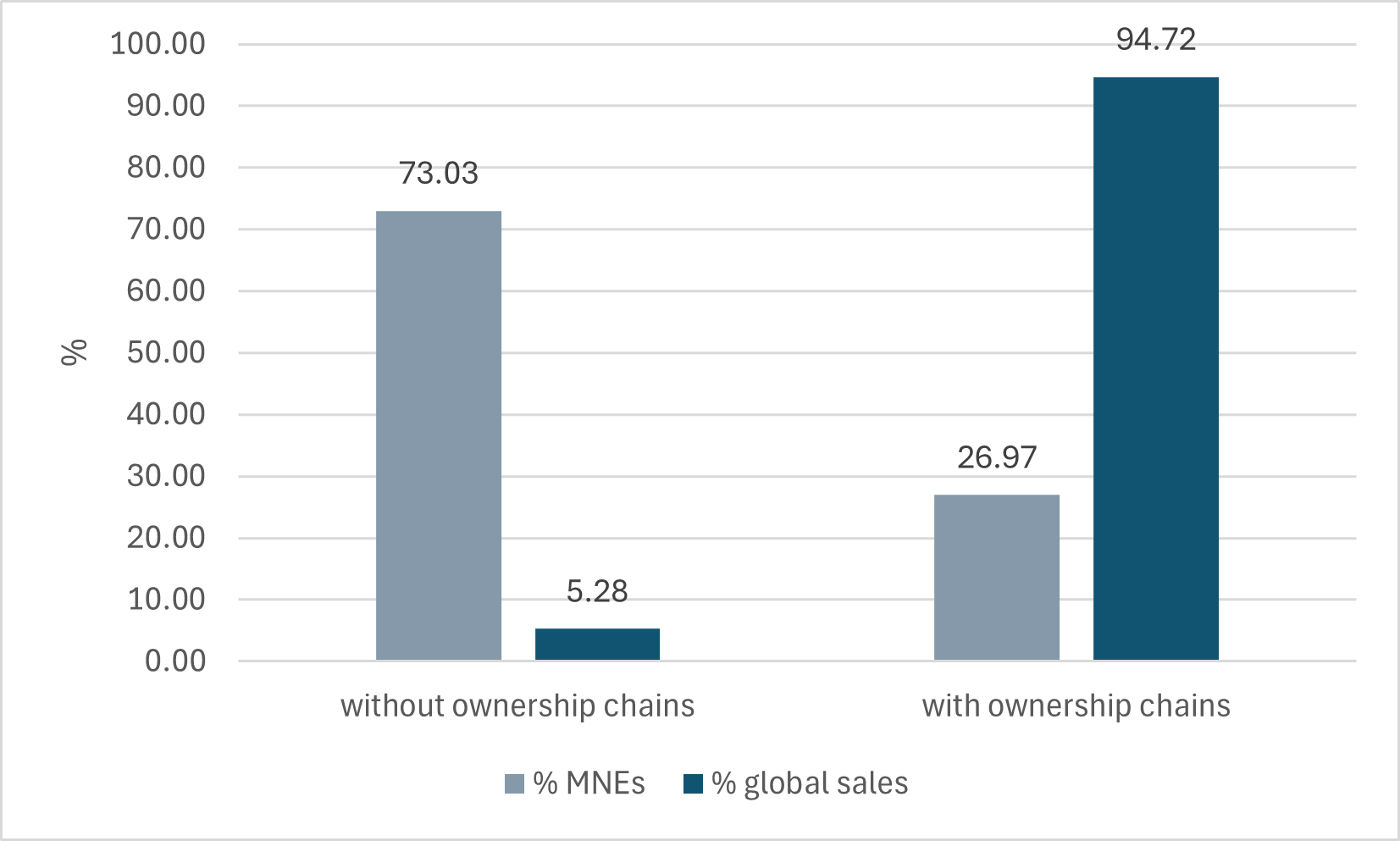}
\begin{tablenotes}
      \footnotesize
      \item Note: On the left, we represent MNEs that have only direct control of subsidiaries, hence no ownership chain. On the right, we represent MNEs with ownership chains. The first bars represent percentages of the total sample by number of parent companies. The second series represents sums of operating revenues generated by the subsidiaries and their parent companies in 2019. 
      \end{tablenotes}
\end{figure}

\noindent Complex multinational enterprises that operate through ownership chains account for a disproportionate share of global economic activity, accounting for 95\% of total MNE revenues, despite representing only 27\% of the firms in our dataset. In Figure \ref{fig: economic weight}, we distinguish between MNEs that exclusively use direct control over subsidiaries (left panel) and MNEs that include at least one ownership chain (right panel). While the latter group is numerically smaller, it dominates global sales, reinforcing the idea that larger corporate networks tend to be structured through multi-tiered ownership chains.\footnote{Our findings are consistent with previous studies by \cite{rungi2017global, altomonte2013business}, which show that the largest percentile of MNEs by corporate perimeter accounts for approximately 75\% of global sales.} This pattern emerges because MNEs without ownership chains tend to be smaller in size. On average, they generate 0.107 billion USD in revenue, compared with 1.782 billion USD for complex MNEs with chains of ownership.\footnote{Ownership data tend to be more comprehensive than financial accounts because tax authorities require full disclosure of shareholding structures, while financial reporting requirements vary across jurisdictions. As a result, financial data may be incomplete due to exemptions and reporting discrepancies. In our analysis, we assume that missing sales data are not systematically correlated with firm size. Alternative imputation methods do not significantly alter our results, as confirmed in Table \ref{fig: economic weight}. Appendix Figure \ref{fig: nonmiss_sh} illustrates that sample selection bias is often driven by geographic coverage rather than firm size.}

\begin{itemize}
    \item \textsl{MNEs are heterogeneous by number of subsidiaries, number of ownership chains, number of countries in which they locate subsidiaries, and number of industries in which subsidiaries operate.} 
\end{itemize}

\begin{figure}[H]
\centering
\caption{MNEs' corporate perimeters}
\label{fig: group size}
\includegraphics[width=.7\linewidth]{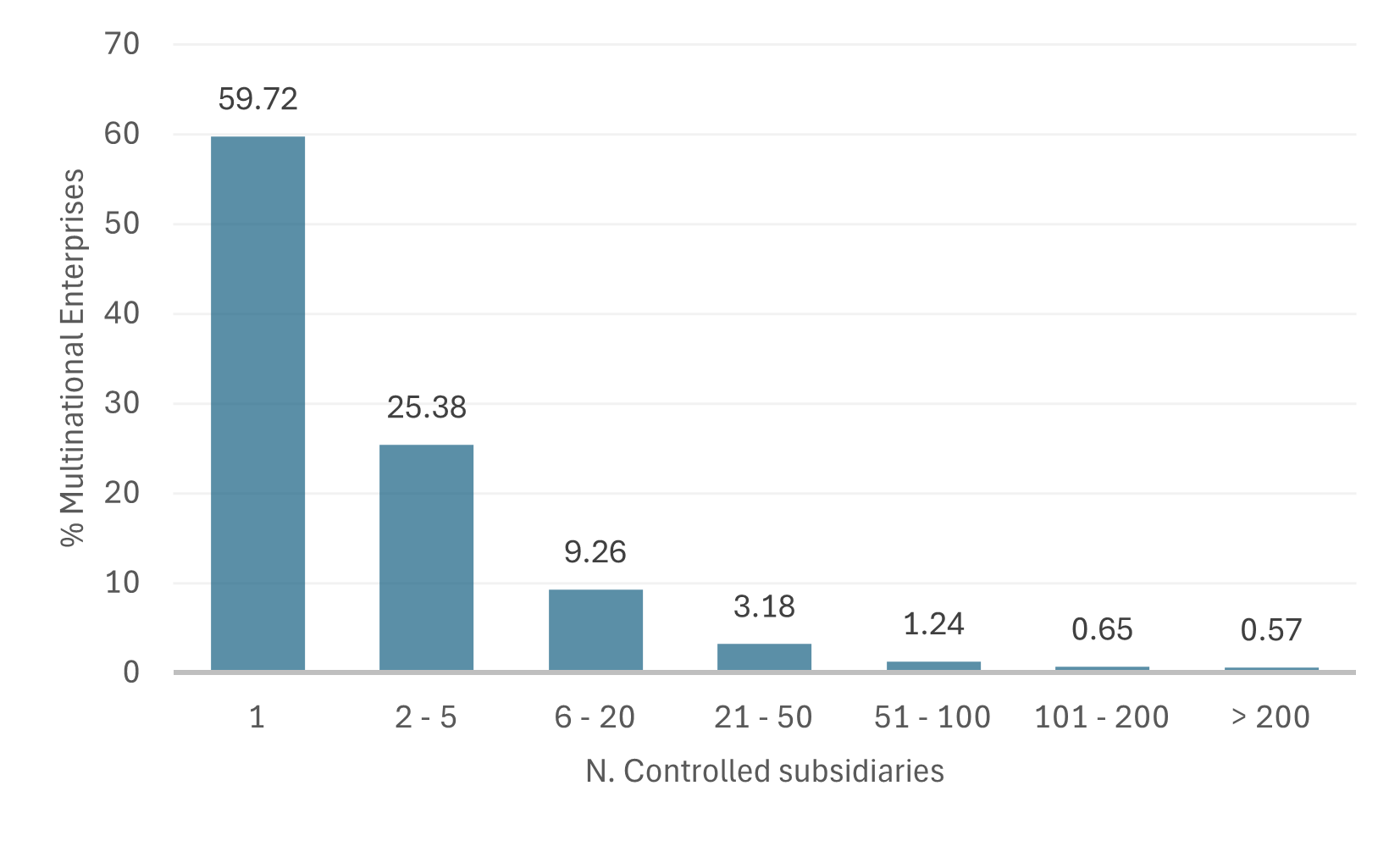}
\begin{tablenotes}
      \scriptsize 
      \item Note: Number of controlled subsidiaries per MNE on the x-axis and the frequency of MNEs on the y-axis. The first bar indicates the simplest MNEs, consisting of one parent and one subsidiary. Complexity of ownership structure is possible only with more than two subsidiaries. 
      \end{tablenotes}
\end{figure}

MNEs exhibit considerable heterogeneity in their organizational structure, varying in the number of subsidiaries, ownership chains, host countries, and industries in which they operate. In Figure \ref{fig: group size}, we classify MNEs according to the number of subsidiaries they control (x-axis) and report their frequency distribution (y-axis). The first bar represents the simplest MNEs, which control only one subsidiary. More complex MNEs, owning at least two subsidiaries, can choose to structure them along an ownership chain. The distribution is highly skewed, with a long right tail, indicating that only 1.2\% of MNEs control more than 100 subsidiaries. This heterogeneity goes beyond the number of subsidiaries controlled. MNEs also vary significantly in the number of countries in which they operate and the diversity of industries in which their subsidiaries operate, further highlighting the complexity of global corporate structures.

\noindent Table \ref{tab: skewness} provides further evidence of the heterogeneity of MNEs' corporate structures. Looking first at the number of subsidiaries controlled, we find an average of 7.93 subsidiaries per parent company. But this masks a highly skewed distribution. The median MNE controls only one subsidiary, while the third quartile of MNEs reaches three subsidiaries per parent company. Moving to the right tail, we find that MNEs at the 90th percentile have more than 10 subsidiaries, and only at the 99th percentile do we reach the most complex MNEs, which control more than 120 subsidiaries, with a maximum of 5,075 subsidiaries worldwide.\footnote{This is the case of Berkshire Hathaway, the largest multinational conglomerate holding company in the United States, ultimately owned by Warren Buffett.} \footnote{In Appendix Figure \ref{fig: size density}, we estimate the distribution of controlled subsidiaries using a negative binomial regression, which yields a mean of 1.882 and an overdispersion parameter of 0.581. The figure visually represents a log-log plot of the frequency distribution of subsidiaries per parent.} A similarly skewed pattern is observed when focusing on intermediate affiliates. The median MNE has zero intermediate affiliate, but the right tail expands sharply, with a jump from 2 to 20 subsidiaries between the 90th and 99th percentiles.\\

\noindent Looking at geographic dispersion, Table \ref{tab: skewness} shows that the average MNE operates in 2.81 countries. Since our definition of an MNE requires at least one foreign subsidiary, the minimum number of countries is two. However, the 90th percentile includes firms operating in four countries, while at the 99th percentile the number jumps to 19 countries and reaches a maximum of 147.\footnote{This extreme case corresponds to DHL, the globally recognized logistics and courier group.} Finally, industry diversification follows a similar pattern. The most complex MNEs, those in the right tail of the distribution, operate in more than 15 industries at the 99th percentile, reaching a maximum of 64 industries.\footnote{According to our data, this is the case of HSBC Holdings PLC, originally founded as the Hongkong and Shanghai Banking Corporation and now a British banking group headquartered in London.}

\begin{table}[H]
    \centering
    \caption{Heterogeneous distributions of MNEs}
        \resizebox{.7\textwidth}{!}{%
    \begin{tabular}{lcccccccc}
    \hline
        Variable & Mean & st. dev. & Min & p50 & p75 & p90 & p99 & Max \\ \hline
        ~ & ~ & ~ & ~ & ~ & ~ & ~ & ~ & ~ \\ 
        N. controlled subsidiaries & 7.93 & 50.09 & 1 & 1 & 3 & 10 & 120 & 5,075 \\ 
        ~ & ~ & ~ & ~ & ~ & ~ & ~ & ~ & ~ \\ 
        N. middlemen & 1.18 & 7.72 & 0 & 0 & 1 & 2 & 20 & 688 \\ 
        ~ & ~ & ~ & ~ & ~ & ~ & ~ & ~ & ~ \\ 
        N. of industries & 2.19 & 3.01 & 1 & 1 & 2 & 4 & 15 & 64 \\ 
        ~ & ~ & ~ & ~ & ~ & ~ & ~ & ~ & ~ \\ 
        N. of countries & 2.81 & 3.92 & 2 & 2 & 2 & 4 & 19 & 147 \\ \hline
    \end{tabular}}
    \label{tab: skewness}
    \begin{tablenotes}
    \footnotesize
        \item Note: The table reports moments of distributions of MNEs. Industries are considered at the 2-digit NACE rev. 2 level. Countries and territories follow the ISO code 2-digit classification. 
    \end{tablenotes}
        \end{table}

\begin{itemize}
    \item \textsl{Bigger multinational enterprises have more subsidiaries and, on average, longer ownership chains.} 
\end{itemize}

\begin{figure}[H]
\centering
\caption{Number of controlled subsidiaries and maximum ownership chain length}
\label{fig: size and distance}
\includegraphics[width=.7\linewidth]{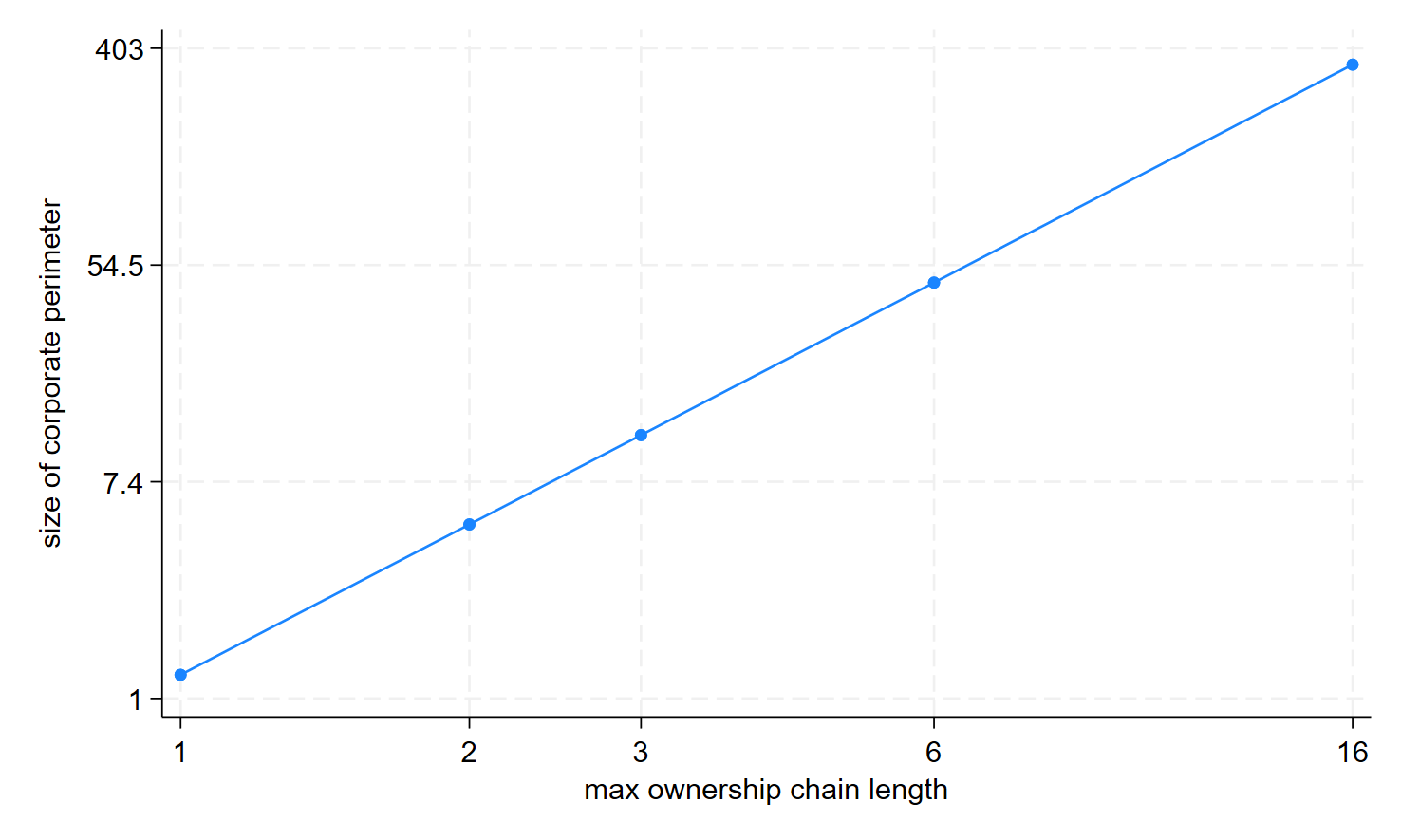}
\begin{tablenotes}
     \footnotesize
     \item Note: The figure shows predicted margins after a log-log regression model where the dependent variable is the number of controlled subsidiaries, the regressor is the maximum length of ownership chains by MNE, and whose coefficient indicates an elasticity equal to 2.01.
\end{tablenotes}
\end{figure}

\noindent The relationship between the number of subsidiaries controlled by an MNE and the maximum length of its ownership chains is illustrated in Figure \ref{fig: size and distance}. The plot represents the predicted ranges of a regression model in which the dependent variable is the logarithm of the firm's scope and the maximum chain length (also in logs) is the key explanatory variable. The relationship appears to be non-linear, with an estimated elasticity of 2.01, suggesting that ownership chains tend to lengthen disproportionately as firms grow. However, large MNEs do not always adopt deep hierarchical structures. Some firms maintain flat ownership networks in which all subsidiaries remain directly controlled by the parent company.\footnote{For example, Assyce Fotovoltaica SL, a Spanish provider of solar energy equipment, directly controls approximately 300 subsidiaries without intermediate affiliates.}

\end{document}